\documentclass[aps,prl,reprint,superscriptaddress,letterpaper,twocolumn,floatfix]{revtex4-2}
\usepackage{amsmath,amssymb,mathtools,dsfont,esint,textgreek,soul}

\usepackage{graphicx,setspace}
\usepackage[usenames, dvipsnames, x11names]{xcolor}
\usepackage[colorlinks]{hyperref}
\usepackage{CJKutf8}
\usepackage{blindtext}
\usepackage{physics}
\usepackage[capitalise]{cleveref}

\begin{document}
\title{Hidden time-reversal in driven XXZ spin chains: exact solutions and new dissipative phase transitions}
\author{Mingxing Yao}
\email{mxyao@uchicago.edu}
\affiliation{Pritzker School of Molecular Engineering, University of Chicago, Chicago, IL 60637, USA}
\author{Andrew Lingenfelter}
\affiliation{Pritzker School of Molecular Engineering, University of Chicago, Chicago, IL 60637, USA}
\affiliation{Department of Physics, University of Chicago, Chicago, IL 60637, USA}
\author{Ron Belyansky}
\affiliation{Pritzker School of Molecular Engineering, University of Chicago, Chicago, IL 60637, USA}
\author{David Roberts}
\affiliation{Joint Quantum Institute, NIST/University of Maryland, College Park, MD, 20742, USA} 
\author{Aashish A. Clerk}
\email{aaclerk@uchicago.edu}
\affiliation{Pritzker School of Molecular Engineering, University of Chicago, Chicago, IL 60637, USA}

\newcommand{\ua}{\uparrow}
\newcommand{\da}{\downarrow}

\newcommand{\al}[1]{\textcolor{blue}{AL: \emph{#1}}}
\newcommand{\aash}[1]{\textcolor{red}{AC: \emph{#1}}}

\newcommand{\jy}[1]{\textcolor{purple}{JY: \emph{#1}}}
\newcommand{\newtext}[1]{\textcolor{purple}{\emph{#1}}}

\begin{abstract}
We show that several models of interacting XXZ spin chains subject to boundary driving and dissipation possess a subtle kind of time-reversal symmetry, making their steady states exactly solvable.  We focus on a model with a coherent boundary drive, showing that it exhibits a unique continuous dissipative phase transition as a function of the boundary drive amplitude.  This transition has no analogue in the bulk closed system, or in incoherently driven models.  We also show the steady state magnetization exhibits a surprising fractal dependence on interaction strength, something previously associated with less easily measured infinite-temperature transport quantities (the Drude weight). Our exact solution also directly yields driven-dissipative double-chain models that have pure, entangled steady states that are also current carrying.  
\end{abstract}

\maketitle

{\it Introduction.---}  
The non-equilibrium dynamics of driven interacting systems represents a forefront in quantum many-body physics.  Among such systems, driven integrable spin chains are a widely studied, paradigmatic class of models \cite{Bertini2021-px,Landi2022-ul,Ilievski2021-iu}.
Interest here is motivated both by their broad relevance to experimental platforms \cite{Mi2024-yp,Wei2022-yj,Rosenberg2024-wj,Boll-2016-Science,Scherg-2018-prl}, and the deep theoretical insights enabled by the tools of quantum integrability.  They exhibit many surprising phenomena, including unexpected connections to classical KPZ universality \cite{Ljubotina2017-jg,Ilievski2018-hc,Ye2022-fa,Ljubotina2019-ga}, transitions between ballistic, diffusive and insulating transport regimes \cite{Prosen2011-kr,Znidaric2011-xm}, and surprising fractal structures in the Drude weight characterizing ballistic transport \cite{Prosen2013-eh,Prosen2011-uf,Ilievski2017-ir}.  Models where boundary driving is balanced with boundary dissipation are also uniquely interesting: their non-equilibrium steady-states (NESS) exhibit a rich set of phenomena that reflect the physics of the bulk Hamiltonian.  Remarkably, Prosen and collaborators have shown that a number of such models (where the driving is incoherent) admit an analytic solution for the NESS \cite{Prosen2011-uf,Prosen_2013_NJP,Prosen2014-od,Prosen2015-vk}.

Given this wide body of work, there are two natural questions that arise. First, it is possible to have phase transitions in the NESS of boundary driven-dissipative spin chains that are controlled by the boundary drive alone, and that {\it do not} simply reflect different phases of the bulk Hamiltonian?  While previous studies have seen transitions in either transport properties or correlation structure \cite{Prosen2008-oc}, they involve tuning a parameter in the bulk Hamiltonian, and are simply related to its structure.  
Second, is there a more general way to understand the remarkable exact solutions in Refs.~\cite{Prosen2011-uf,Prosen2011-kr,Karevski2013-ju,Prosen2015-vk}.
If so, can we use this to derive qualitatively new exact solutions?

\begin{figure}[t]
    \centering
    \includegraphics{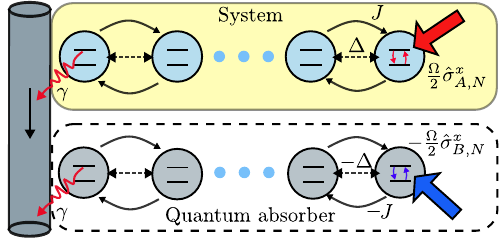}
    \caption{Schematic of the boundary driven XXZ spin chain and the CQA construction of the doubled system.
    The system (top) consists of $N$ spins with a coherent drive at site $N$ (right) and incoherent loss at site $1$ (left). The absorber (bottom) 
    consists of mirroring the system and is coupled to the system via a  unidirectional waveguide. 
    }
    \label{fig:schematic}
\end{figure}

In this work, we address both these questions.  We show that a version of quantum detailed balance (``hidden time reversal" symmetry (hTRS) \cite{Roberts2021-cz}) is present in a variety of boundary driven-dissipative XXZ spin chain models, enabling exact analytic solutions of the NESS via the coherent quantum absorber construction  \cite{Stannigel2012-kk,Roberts2021-cz} (see  \cref{fig:schematic}).  This provides a new, physically-appealing way to understand previously-derived exact solutions, and lets us derive completely new solutions.  We use this to find an exact solution for the NESS of a XXZ spin chain subject to boundary loss and a {\it coherent} boundary drive (see Fig. \ref{fig:schematic}).
Dissipative many-body systems with coherent boundary drives are relatively unstudied, but can have remarkable behaviour \cite{Fitzpatrick2017-xe,Prem2023-ab}.  Unlike previously studied XXZ models with incoherent drives, our NESS exhibits a continuous phase transition as a function of the boundary drive amplitude, a transition that has no simple correspondence to ground-state phases of the bulk XXZ Hamiltonian.  The transition can be seen by simply measuring the average magnetization or spin current.  We also find that the steady-state magnetization mirrors the the surprising fractal structure of the infinite-temperature XXZ model Drude weight \cite{Prosen2013-eh}, an effect related to the quasiparticle structure of the bulk XXZ Hamiltonian. 
This represents perhaps the simplest observable manifestation of this effect.

Our work reveals other surprising structures.  The connection to hTRS allows us to also construct exactly-solvable models where two boundary-driven XXZ chains are coupled via common waveguide (see Fig.~\ref{fig:schematic}).  These systems exhibit a pure, current-carrying steady state, with a steady state entanglement that grows logarithmically with system size.  Further, the realization that these driven-dissipative interacting spin models have hTRS does more than let us obtain the steady state: it also implies that a certain class of steady-state correlation functions exhibit Onsager symmetry.


{\it Model.---}We consider a driven-dissipative system of $N$ spin-1/2s governed by the GKSL (Lindblad) master equation
\begin{equation}
    \partial_t \hat{\rho} = \hat{\mathcal{L}} \hat{\rho} = -i[\hat{H}_{\rm XXZ}+ \frac{\Omega}{2} \hat{\sigma}_N^x, \hat{\rho}] + \gamma \mathcal{D}[\hat{\sigma}_1^-] \hat{\rho}, \label{eq:me-coherent}
\end{equation}
where $\hat{H}_{\rm XXZ} = \sum_{j = 1}^{N-1} (J \hat{\sigma}_j^x  \hat{\sigma}_{j+1}^x + J \hat{\sigma}_j^y  \hat{\sigma}_{j+1}^y  + \Delta \hat{\sigma}_j^z\hat{\sigma}_{j+1}^z)$ is the XXZ Hamiltonian, with $\hat{\sigma}^x,\hat{\sigma}^y,\hat{\sigma}^z$ being the Pauli matrices, $\Omega$ is the amplitude of the coherent drive on site $N$, and $\mathcal{D}[\hat{L}] \hat{\rho} \equiv \hat{L}\hat{\rho} \hat{L}^\dagger - \frac{1}{2} \{\hat{L}^\dagger \hat{L},\hat{\rho} \}$ is the Lindblad dissipator, describing decay from boundary site one with jump operator $\hat{L}=\hat{\sigma}_1^-$ and rate $\gamma$.  Eq.~(\ref{eq:me-coherent}) describes a 1D chain of qubits with nearest neighbor hopping and $ZZ$ interactions, with a Rabi drive on site $N$ (treated in the rotating-wave approximation) and loss on site $1$. This setup is naturally feasible in experimental platforms such as superconducting circuits and Rydberg atoms \cite{Fitzpatrick2017-xe,Fedorov2021-cr,Kounalakis-2018-njp,Scholl-2022-prxq}.

In this work, we focus on the exact steady state $\hat{\rho}_{\rm ss}\equiv \lim_{t\rightarrow \infty}e^{\hat{\mathcal{L}}t}\hat{\rho}(0)$ of this master equation, which, as we show in the Supplemental Material (SM) \cite{supp}, is unique for all parameters $J,\,\Delta,\,\Omega,\,\gamma$.

\begin{figure*}
    \centering
    \includegraphics[width = \textwidth]{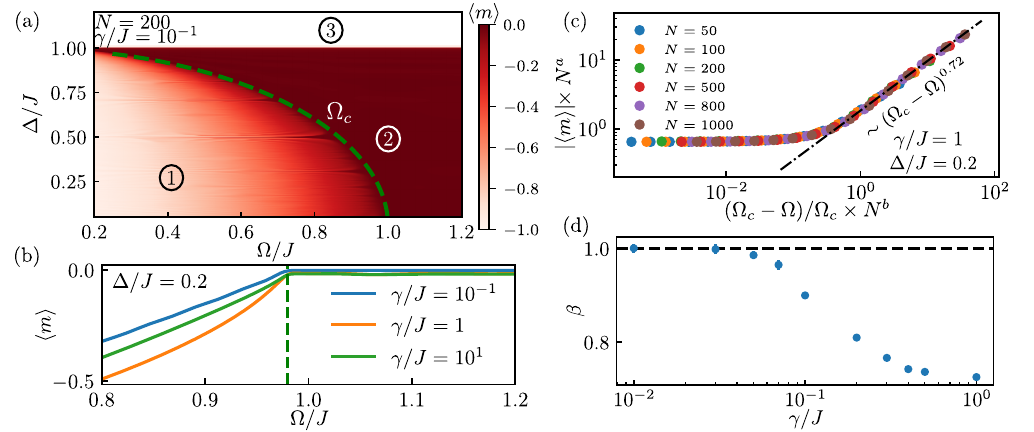}
    \caption{Phase diagram. (a) Magnetization density $\langle m \rangle$ in the steady state as a function of $\Delta/J$ and $\Omega/J$. The three phases are denoted by $(1)$ the ballistic ordered phase, $(2)$ ballistic disordered phase, $(3)$ insulating ordered phase.  The dashed green line denotes the analytically found phase boundary \cref{eq:Omega-crit}. (b) A line cut of the magnetization density for fixed $\Delta/J = 0.2$ and system size $N=500$. (c) Finite size scaling of the magnetization near the critical drive $\Omega_c$. The exponents are $a= 0.55$ and $b=0.75$. The scaling exponent agrees with $\beta = a/b$. (d) Critical exponent $\beta$ as a function of $\gamma/J$. For $\gamma/J> 0.1$ we extract the critical exponent by fitting the finite-size scaling function (as in (c)), but for weaker dissipation ($\gamma/J\leq0.1$) we only use the largest available system size due to limitations posed by finite size effects (see SM \cite{supp} for details). Black dashed line represents the analytical prediction $\beta = 1$ in the weak dissipation limit. }
    \label{fig:phase-diagram}
\end{figure*}

We find the steady state of  Eq.~\eqref{eq:me-coherent} using the coherent quantum absorber (CQA) method \cite{Stannigel2012-kk,Roberts2021-cz}.
The basic idea of CQA is to redirect the dissipation of the system of interest (``$A$") to an absorber system (``$B$") via a chiral waveguide (see \cref{fig:schematic}). The absorber is designed such that the combined $A+B$ system relaxes to a pure dark state $|{\psi_{\rm CQA}}\rangle$. The steady state of system $A$ can then be retrieved by tracing out the absorber $\hat{\rho}_{\rm ss} = {\rm Tr}_B (|\psi_{\rm CQA} \rangle \langle \psi_{\rm CQA} |)$.
For a system with hidden time-reversal symmetry \cite{Roberts2021-cz} (as we find to be the case here), there is a simple construction of the absorber system $B$:  one simply mirrors the  original spin chain with coherent dynamics $\hat{H}^{(B)} = - \hat{H}^{(A)}$ and jump operator $\hat{L}^{(B)} = - \hat{L}^{(A)}$ (see SM for details \cite{supp}).
The dynamics of this doubled system $\hat{\rho}_{\rm AB}$ is described by the cascaded master equation $\partial_t \hat{\rho}_{\rm AB} = - i [\hat{H}^{(AB)},\hat{\rho}] + \gamma \mathcal{D}[\hat{L}^{(AB)}] \hat{\rho}$, with the coherent dynamics and the jump operator described by 

\begin{align}
    \hat{H}^{(AB)} & = \hat{H}^{(A)}_{\rm XXZ} - \hat{H}^{(B)}_{\rm XXZ} + \frac{\Omega}{2} (\hat{\sigma}_{A,N}^x - \hat{\sigma}_{B,N}^x ) + \hat{H}_{\rm c}\\
    \hat{L}^{(AB)} & = \hat{\sigma}_{A,1}^- - \hat{\sigma}_{B,1}^-
\end{align}
where $\hat{H}_{\rm c} = -\frac{i\gamma}{2} (\hat{\sigma}_{A,1}^+ \hat{\sigma}_{B,1}^- - \hat{\sigma}_{A,1}^- \hat{\sigma}_{B,1}^+)$ is the chiral coupling between the system and the absorber \cite{Carmichael1993-dm,Gardiner1993_Cascaded}. We construct the CQA state by searching for pure state solution of the dark state conditions $\hat{H}^{(AB)} \ket{\psi_{\rm CQA}} = 0$ and $\hat{L}^{(AB)} \ket{\psi_{\rm CQA }} = 0$.

We consider the doubled system to be a chain of $N$ sites, but with a doubled local Hilbert space spanned by the four states: 
$|0_i\rangle\equiv  {|{\da_{A,i} \da_{B,i}\rangle}} $, $|1_i\rangle\equiv  {|{\ua_{A,i} \ua_{B,i}}\rangle} $, $|S_i\rangle\equiv  ({|{\ua_{A,i} \da _{B,i}}\rangle} - {|{\da_{A,i} \ua_{B,i}}\rangle})/\sqrt{2}$, and $|T_i\rangle\equiv  ({|{\ua_{A,i} \da _{B,i}}\rangle} + {|{\da_{A,i} \ua _{B,i}}\rangle})/\sqrt{2}$.  Here
${|{\ua} \rangle}$, ${|{\da} \rangle}$ denote $\hat \sigma_z$ eigenstates with eigenvalues $\pm1$, respectively, and $\hat \sigma^- = {|{\da}\rangle\langle{\ua}|}$.
Inspired by the Matrix Product Operator (MPO) ansatz of Ref.~\cite{Prosen2011-uf}, we construct the pure (unnormalized) dark state as a Matrix Product State (MPS) ansatz
\begin{equation}
    |\psi_{\rm CQA}\rangle = \sum_{s_1,\cdots, s_N} \vb{v_L^\dagger} \hat{A}_{s_1} \cdots \hat{A}_{s_N} \vb{v_R} |s_1 \cdots s_N \rangle \label{eq:solution}.
\end{equation}
Here, the summation is limited to the triplet subspace $s_i \in \{0_i,1_i,T_i\}$, as the singlet state $|S_i\rangle$ decouples due to the dark state condition with zero energy $E = 0$ (see SM \cite{supp}). 
The matrices $\hat{A}_s$ are taken to be translationally invariant and represented by the following ansatz: 
$\hat{A}_T = \sum_{k=0}^{N} a_k |k\rangle \langle k|$, $
\hat{A}_1 = \sum_{k=0}^{N-1} b_k |k+1\rangle \langle k|$, and $\hat{A}_0 = \sum_{k=0}^{N-1} c_k |k\rangle \langle k+1|$, where $\{|k\rangle, k = 0,1, \cdots,N\}$ labels the auxiliary space (which can be thought of as an auxiliary one-dimensional lattice). The left vector $\vb{v_L}$, corresponding to the dissipative site, can be chosen to be simply $|0\rangle$ (i.e.~the leftmost site in the auxiliary lattice), while the right vector $\vb{v_R}$, corresponding to the coherently driven site, is parameterized as $\vb{v_R} = \sum_{k=0}^{N} \alpha_k |k\rangle$.
Imposing the dark state condition on the state in \cref{eq:solution} then yields recursion relations for the matrix elements $a_k,b_k,c_k$ as well as for the coefficients $\alpha_k$.
These relations can be efficiently solved numerically exactly but also approximately analytically in certain limits, shown in the SM \cite{supp}.
Note that $\hat{A}_s$ changes the auxiliary state by at most one, which implies that the bond dimension of the MPS scales at most linearly with $N$.

The $\hat{A}_s$ matrices are identical to those appearing in the MPO solution of previous studies of incoherently pumped XXZ spin chains \cite{Prosen2011-uf,Karevski2013-ju} and are fully determined by the dissipation $\gamma$ and bulk parameters $J$ and $\Delta$.
Unlike those studies, however, where the right vector
$\vb{v_R}$ has a trivial form determined solely by the form of the Lindbladian on the physical boundary site, here it has a more complex structure: it encodes the interplay between the bulk Hamiltonian and the coherent boundary drive. 
As we show, this non-trivial competition leads to a phase transition with no analogue in incoherently pumped models.
More generally, the CQA construction employed in this work allows us to generalize the results of Refs.~\cite{Prosen2011-uf,Karevski2013-ju}, providing a pure steady state of the corresponding doubled system (see the SM \cite{supp} for an example).

{\it Phase diagram.---}
From the exact solution, \cref{eq:solution}, we can efficiently determine the phase diagram of the model, shown in \cref{fig:phase-diagram}(a) in the $\Omega, \Delta$ plane. We use the magnetization density $\langle {m}\rangle \equiv \sum_{i=1}^N {\rm Tr} (\hat{\sigma}_i^z \hat{\rho}_{\rm ss})/N$ as an order parameter, which distinguishes three different regimes, marked in \cref{fig:phase-diagram}(a) as $(1)$, $(2)$, and $(3)$.
One phase boundary shows up along the line $\Delta/J=1$, corresponding to a Heisenberg exchange interaction.  For $\Delta > J$, the system is always ordered irrespective of the drive $\Omega$. This transition corresponds to the ground state phase transition of the bulk XXZ spin chain between a Luttinger liquid with ferromagnetic order and a paramagnetic phase \cite{Franchini2016-fh}.
More interesting is the regime $\Delta/J < 1$, where we find an additional, genuinely non-equilibrium, transition that has no analog in the bulk closed system. 

For $\Delta<J$, we find a continuous transition as the coherent drive $\Omega$ is increased from zero, from an ordered state [region (1) with $\langle m \rangle \neq 0$] to a disordered state [region (2) with $\langle m \rangle =0$] at a critical drive $\Omega_c$ [green dashed line in \cref{fig:phase-diagram}(a)]. 
The transition can be understood from a quantum-to-classical mapping of the MPS \cref{eq:solution}, under a suitable gauge choice.
Local observables, such as the magnetization, can be equivalently calculated from a classical stochastic process consisting of a particle hopping on a chain of $N$ sites.
In the weak dissipation limit, $\gamma/J \to 0$ 
\footnote{As in many driven dissipative systems, the order of taking the weak dissipation limit $\gamma/J \to 0$ and the thermodynamic limit $N\to \infty$ do not commute. Here we first take the thermodynamic limit and then investigate the system in the vanishing dissipation limit.}, the hopping rates (given by the coefficients of the $\hat{A}$ matrices) can be analytically obtained and manifest as quasiperiodically disordered rates.
Within this classical model, the magnetization corresponds to the average distance traveled in $N$ steps starting from the state $\vb{v_L}$ (i.e., site zero), with the final position weighted by the probability distribution corresponding to $\vb{v_R}$ (something that can be interpreted as a post-selection).
The phase transition can be then be understood as a localization transition of the right vector $\vb{v_R}$.
In the weak dissipation limit $\gamma/J \to 0$, the coefficient $\alpha_k$ is given by a Chebyshev polynomial of the first kind $\alpha_k(\Omega_c/\Omega) = T_{k}(\Omega_c/\Omega)$, where
\cite{supp}
    \begin{equation}
    \label{eq:Omega-crit}
        \Omega_c  =\sqrt{J^2 - \Delta^2}.
\end{equation}
The phase transition occurs at the point $\Omega_c/\Omega = 1$ where $\vb{v_R}$ transitions from exponentially localized $\alpha_k \sim e^{\xi k}$ with $\cosh \xi = \Omega_c/\Omega$ (for $\Omega <\Omega_c$), which favors trajectories with large excursions away from the initial site (zero), giving rise to $\langle m\rangle \approx -1$, to oscillatory $\alpha_k \sim \cos k \xi $ with $\cos \xi = \Omega_c/\Omega$ (for $\Omega >\Omega_c$), which leads to a disordered state $\langle m \rangle = 0$.

While \cref{eq:Omega-crit} is derived from the weak dissipation limit, we numerically find that the critical drive holds approximately for arbitrary dissipation strength [see Fig.~\ref{fig:phase-diagram}(b) and SM \cite{supp}].
Although the critical drive $\Omega_c$ does not directly control ground state properties of the bulk closed system, it surprisingly is proportional to the square root of the smooth part of the infinite temperature spin Drude weight $D$ \cite{Prosen2013-eh}, which characterizes the ballistic contribution of the real part of the spin conductivity.   
Further, the Drude weight can be interpreted as an effective velocity squared \cite{Nardis-2019-scipost}.  Our phase transition thus can be interpreted  as a competition between the drive and an effective propagation velocity.  This is  reminiscent of the phase transition in the asymmetric simple exclusion process \cite{Derrida1998-ti}, which involves an analogous competition between a drive strength and a hopping rate.  

To better understand the phase transition, we now turn to the critical behaviour near the critical point, where $|{\langle {m} \rangle}| \sim (1- \frac{\Omega}{\Omega_c})^\beta$ for $\Omega<\Omega_c$. We extract the exponent $\beta$ by a finite size scaling, shown in Fig.~\ref{fig:phase-diagram}(c).
Surprisingly, while the critical drive is independent of dissipation, we find that the critical exponent $\beta$ varies continuously as a function of $\gamma$, as shown in Fig.~\ref{fig:phase-diagram}(d). In the $\gamma/J \rightarrow 0$ limit, we analytically find the exponent $\beta=1$ (see SM \cite{supp}), which agrees well with the numerically extracted exponents in that limit [Fig.~\ref{fig:phase-diagram}(d)].

\begin{figure}
    \centering
    \includegraphics{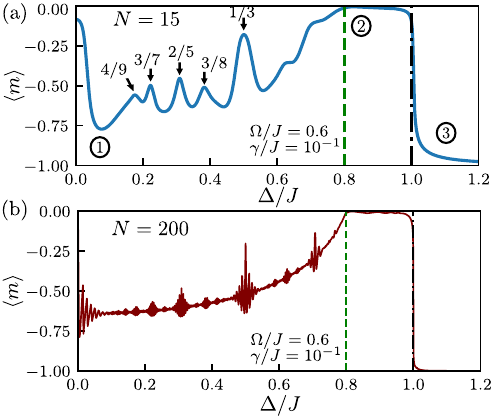}
    \caption{The steady state magnetization as a function of $\Delta/J$ (a) $N=15$ and (b) $N=200$.
    The green dashed line corresponds to the critical point \cref{eq:Omega-crit} separating phases (1) and (2) and the black dashed line ($\Delta=1$) separates phases (2) and (3). 
    In (a), we can identify all the resonances due to the special points $l/m$ in \cref{eq:special-Delta}.
    }
    \label{fig:fractal}
\end{figure}

Another interesting feature in the phase diagram in \cref{fig:phase-diagram}(a) for $\Delta/J <1$ is the presence of a set of interaction parameters $\Delta$ where the system shows strikingly different behaviour. These occur when $\Delta$ is tuned to
    \begin{equation}
    \label{eq:special-Delta}
        \Delta_{l,m} /J = \cos \frac{l}{m} \pi, \quad m,l\in \mathbb{Z}.
    \end{equation}
For these $\Delta$, 
and in the weak-$\gamma$ limit, one can rigorously show that
the bond dimension of the exact solution MPS in \cref{eq:solution} becomes finite (i.e.~independent of system size).
 These special points are well known in another context, namely studies of the fractal structure of the spin Drude weight in the infinite temperature XXZ model (see e.g.~\cite{Prosen2011-kr,Prosen2013-eh, Ilievski2017-ir,VasseurPRB2020,Bulchandani2021-rq}).
Here, we find the magnetization versus $\Delta$ exhibits resonances peaks near each $\Delta_{l,m}$ [see Fig.~\ref{fig:fractal}], and further, there is no phase transition when $\Delta$ is exactly tuned to 
$\Delta_{l,m}$ (see SM \cite{supp}). 
Note that the free fermion case is a special point corresponding to $\Delta_{1,2}=0$.
Our model thus presents a direct manifestations of this fractal structure, showing up in an observable (the steady-state magnetization) that is far more experimentally accessible than the Drude weight.   
Note that the number of observable resonance peaks is constrained by system size, and that their resonance width diminishes with increasing $N$ [See \cref{fig:fractal}(b)]. As we discuss in the SM \cite{supp}, this permits the phase transition to persist in the thermodynamic limit for irrational $\Delta$.

{\it Doubled system and hTRS.---}
We now turn to discussing several interesting features of the doubled system of \cref{fig:schematic} and consequences of the hTRS. We first note that the different phases in \cref{fig:phase-diagram} can also be distinguished by the spin current $\langle j \rangle \equiv {i \langle (\hat{\sigma}_{A,m}^+ \hat{\sigma}_{A,m+1}^- - {\rm H.c.}) \rangle_{\rm ss}}$. Specifically, phase (3) is insulating, with $\log \expval{j}\propto - N$, whereas both phases (1) and (2) are ballistic (with the coefficient showing nonanalytic behaviour similar to the magnetization in \cref{fig:phase-diagram}(b); see SM \cite{supp}). Exactly at $\Delta/J = 1$, the current is subdiffusive $\expval{j}\sim N^{-2}$, similar to previous studies on XXZ spin chains \cite{Landi2022-ul,Prosen2011-kr}. 
Moreover, we note that the current can also be viewed as the steady-state current of the doubled system in \cref{fig:schematic}, travelling from the coherent drive on site $N$ of the system, via the chiral waveguide, through the absorber and exiting via the coherent drive on site $N$ of the absorber system. This implies an exotic situation where 
the cascaded master equation stabilizes a pure, current carrying state $|\psi_{\rm CQA}\rangle$.  We stress that there are a variety of means of realizing such effective chiral waveguides in experiments~\cite{Cao2024-vz}.

\begin{figure}
    \centering
    \includegraphics{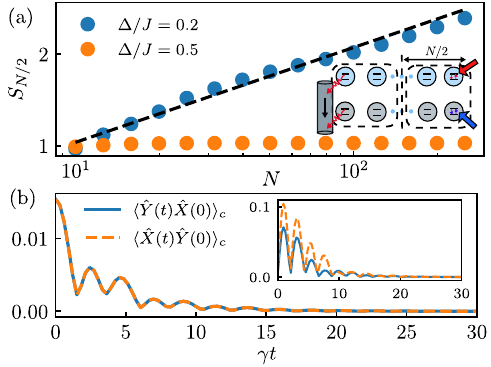}
    \caption{Features of double construction and hTRS (a) Entropy scaling for general anisotropy $\Delta/J = 0.2$ and special points $\Delta/J = \cos \frac{\pi}{3}=0.5$ with $\Omega/J = 2$ and $\gamma/J = 1$. The stark difference on scaling implies the effective system size. (Inset: the bipartition for entropy) (b) hTRS: two time correlators for a small single chain system $N=3$ with parameters $\Omega/J = 0.4$, $\Delta/J =0.2$, and $\gamma/J = 1$. It shows symmetric correlation between $\hat{X} = \hat{\sigma}^-_1$ and $\hat{Y} = \hat{\sigma}^-_1 \hat{\sigma}^-_2$. This symmetry does not hold true for general spin operators (inset: $\hat{X} = \hat{\sigma}_1^z $ and  $\hat{Y} = \hat{\sigma}_2^z$).}
    \label{fig:h-TRS}
\end{figure}

We can also study quantities in this doubled system that do not correspond to quantities in the single spin chain.  For example, using the pure steady state of \cref{eq:solution}, we can look at the von Neumann entanglement entropy for a bipartition corresponding to a vertical cut in Fig.~\ref{fig:schematic}:  $S_{N/2} = - {\rm Tr} (\hat{\rho}_{N/2} \log \hat{\rho}_{N/2})$, where $\hat{\rho}_{N/2} = {\rm Tr}_{x>N/2} (|\psi_{\rm CQA} \rangle \langle \psi_{\rm CQA }|)$ is the reduced density matrix for half (all sites $x>N/2$) of the combined system-absorber. 
As discussed in the previous section, at the special values of $\Delta$ in \cref{eq:special-Delta}, the MPS in \cref{eq:solution} has a finite bond dimension, implying area law entanglement, which we confirm numerically in \cref{fig:h-TRS}(a). On the other hand, away from the special points for $0<\Delta/J < 1$, we find logarithmic scaling $S_{N/2} \sim \log N$, implying that that linear growth of the bond dimension is required for accurate representation of the state \cref{eq:solution}. This is yet another striking difference between special points and general points as it persists in the disordered phase $(2)$ where the resonance structure vanishes.

Finally, we note that the presence of hTRS in our system also puts constraints on its dynamics \cite{Roberts2021-cz}. In particular, it implies the following Onsager symmetry for certain operators \cite{supp}:
\begin{equation}
    {\rm Tr} (\hat{X}(t) \hat{Y}(0) \hat{\rho} _{\rm ss}) = {\rm Tr} (\hat{Y}(t) \hat{X}(0) \hat{\rho} _{\rm ss}). \label{eq:hTRS}
\end{equation}
Here, $\hat{X}$ and $\hat{Y}$ are the time symmetric operators predicted by hTRS \cite{supp}. In the case of Eq.~\eqref{eq:me-coherent}, we found that the following local operators $\hat{X} = \hat{\sigma}_1^- $ and $\hat{Y} = \hat{\sigma}_1^- \hat{\sigma}_2^-$ are time symmetrical, which we verified numerically for a small system in \cref{fig:h-TRS}(b). In contrast, the spin correlation functions $\langle \hat{\sigma}_1^z (t) \hat{\sigma}_2^z(0) \rangle_{\rm ss}$ and $\langle \hat{\sigma}_2^z (t) \hat{\sigma}_1^z (0)\rangle_{\rm ss}$ are not symmetric in time [see the inset in \cref{fig:h-TRS}(b)]. This is related to the fact that the Onsager symmetry cannot hold for every pair of operators, as the current breaks time reversal symmetry explicitly.

{\it Summary and outlook.---}
In this work, we introduced a novel exact solution to the boundary coherent-driven dissipative XXZ spin chain with an explicit doubled state construction. This allowed us to identify a unique continuous dissipative phase transition as a function of the boundary drive amplitude.  We also found that simple observables in the dissipative steady state directly manifest fractal behavior as a function of the anisotropy parameter $\Delta/J$.  Our work suggests that coherently driven dissipative models could be a rich source of new phenomena and insights.  While our work focuses on the phase transition in the thermodynamic limit, intriguing phenomena exist for small system sizes, such as the fractal resonance structure and subtle hTRS symmetry; these are all extremely well suited for near-term quantum simulation experiments.

\begin{acknowledgments}
	{\it Acknowledgments.---}
    We acknowledge helpful discussions with R. Vasseur, M. Maghrebi, T. Jin, and S. Gopalakrishnan.   
    This work was supported by the the Army Research Office under 
    Grant No.~W911NF-23-1- 0077, the National Science Foundation QLCI HQAN (NSF Grant No. 2016136), the Air Force Office of Scientific Research MURI program under Grant No. FA9550-19-1-0399, and the Simons Foundation through a Simons Investigator Award (Grant No. 669487), and partially supported by
    the University of Chicago Materials Research Science and Engineering Center, which is funded by the National Science Foundation under Grant No. DMR-2011854.
\end{acknowledgments}
\bibliography{xxz-transport}

\end{document}


\renewcommand{\theequation}{S\arabic{equation}}
\renewcommand{\thefigure}{S\arabic{figure}}
\renewcommand{\thesection}{\Roman{section}}
\setcounter{equation}{0}
\setcounter{page}{1}

\widetext
\begin{center}
    \textbf{\large Supplemental Material for ``Hidden time-reversal in driven XXZ spin chains: exact solutions and new dissipative phase transitions''}
\end{center}
\tableofcontents

\section{Coherent quantum absorber and exact solution}

\subsection{CQA solution to the coherent driven-dissipative spin chain}
In this section, we derive the exact solution for the coherent boundary driven-dissipative model. The bulk coherent dynamics is described by XXZ Hamiltonian:
\begin{equation}
    \hat{H}_{\rm XXZ} = \sum_{j = 1}^{N-1} (J \hat{\sigma}_j^x  \hat{\sigma}_{j+1}^x + J \hat{\sigma}_j^y  \hat{\sigma}_{j+1}^y  + \Delta \hat{\sigma}_j^z\hat{\sigma}_{j+1}^z).
\end{equation}
The system is driven by a coherent drive $\frac{\Omega}{2}\hat{\sigma}_N^x$ at site $N$ and coupled to zero temperature reservoir with jump operator $\hat{L} = \sqrt{\gamma}\sigma_1^-$ on site $1$. The dynamics is governed by Lindblad master equation
\begin{align}
    \partial_t \hat{\rho} = \hat{\mathcal{L}} \hat{\rho} = -i[\hat{H}_{\rm XXZ}+ \frac{\Omega}{2} \hat{\sigma}_N^x, \hat{\rho}] + \gamma \mathcal{D}[\hat{\sigma}_1^-] \hat{\rho}. \label{eq-app:sysAqme}
\end{align}
We seek the exact steady state solution $\mathcal{L} \hat{\rho}_{\rm ss} = 0$ of this model for all parameters.

We employ the ``coherent quantum absorber" (CQA) method to find the exact solution \cite{Stannigel2012-kk,Roberts2021-cz}.
The basic idea is to channel the dissipation of the system (call it system ``A") via a chiral waveguide to an absorber system (``B"). For a specific choice of absorber, the combined system + absorber relaxes to a pure steady state $|\psi_{\rm CQA}\rangle$ \cite{Stannigel2012-kk}. Moreover, because the system is upstream the absorber, one may trace out the absorber to obtain the A system steady state: $\hat\rho_{\rm ss} = {\rm Tr}_B(\ketbra{\psi_{\rm CQA}}{\psi_{\rm CQA}})$.
A special class of dissipative systems possess a hidden time reversal symmetry (hTRS) and as a consequence have a particularly simple form of absorber: an mirrored copy of the system with the Hamiltonian $\hat H^{(B)} = - \hat H^{(A)}$ \cite{Roberts2021-cz}.
As we show next, we can solve this model using the CQA construction with $\hat{H}^{(B)} = -\hat H^{(A)}$, where $\hat H^{(A)} = \hat H_{\rm XXZ}^{(A)} + (\Omega/2)\hat\sigma^x_{A,N}$ [See Fig.~1 in the main text], hence the model possesses hTRS.  The system A and absorber B are coupled to a common chiral waveguide reservoir that mediates collective dissipation $\hat{L}^{(AB)} = \hat{\sigma}_{A,1}^- - \hat{\sigma}_{B,1}^-$. The chiral waveguide induces an exchange Hamiltonian $\hat{H}_{\rm c} = -\frac{i\gamma}{2} (\hat{\sigma}_{A,1}^+ \hat{\sigma}_{B,1}^- - {\rm h.c.})$ which is required to ensure nonreciprocal dynamics.
The full system and absorber master equation is
\begin{align}
\partial_t \hat{\rho}_{\rm AB} &= \mathcal{L}_{\rm AB} \hat{\rho}_{\rm AB} = - i [\hat{H}^{(A)}_{\rm XXZ} - \hat{H}^{(B)}_{\rm XXZ} + \frac{\Omega}{2} (\hat{\sigma}_{A,N}^x - \hat{\sigma}_{B,N}^x ) + \hat{H}_{\rm c}, \hat{\rho}_{\rm AB}] + \gamma \mathcal{D}[\hat{\sigma}_{A,1}^- - \hat{\sigma}_{B,1}^-] \hat{\rho}_{\rm AB}\nonumber \\
 & = - i[\hat{H}^{(AB)}, \hat{\rho}_{\rm AB}] + \gamma \mathcal{D}[\hat{L}^{(AB)}] \hat{\rho}_{\rm AB}. \label{eq-app:master}
\end{align}
Note that one may trace out the absorber system from this master equation and recover the system A master equation [cf.~Eq.~\eqref{eq-app:sysAqme}].

To find the pure steady state $\ket{\psi_{\rm CQA}} $ of Eq.~\eqref{eq-app:master}, it is sufficient to construct a state that satisfies the following dark state conditions \cite{Kraus2008-ej},
\begin{equation}
    \hat{H}^{(AB)}\ket{\psi_{\rm CQA}}  = 0, \quad \hat{L}^{(AB)} \ket{\psi_{\rm CQA}}  = 0.
    \label{eq-app:dark-condition}
\end{equation}
(In general it is only required that $|\psi_{\rm CQA}\rangle$ is an eigenstate, but for models with hTRS, the eigenvalue is zero \cite{Roberts2021-cz}.)
To construct the steady state satisfying these dark state conditions, we make a matrix product state (MPS) ansatz for the solution following the previous work \cite{Prosen2011-uf}:
\begin{equation}
    |\psi_{\rm CQA}\rangle = \sum_{s_1,\cdots, s_N} \vb{v_L^\dagger} \hat{A}_{s_1} \cdots \hat{A}_{s_N} \vb{v_R} |s_1 \cdots s_N \rangle \label{eq-app:solution}.
\end{equation}
Here, each state $|s_i\rangle$ is a two-spin dimer state of the A spin and B spin of site $j$ and is represented in the triplet-singlet basis ${|0_i\rangle}\equiv  {|{\da_{A,i} \da_{B,i}}\rangle} $, ${|1_i\rangle}\equiv  {|{\ua_{A,i} \ua_{B,i}}\rangle} $, ${|S_i\rangle}\equiv  ({|{\ua_{A,i} \da _{B,i}}\rangle} - {|{\da_{A,i} \ua_{B,i}}\rangle})/\sqrt{2}$, and ${|T_i\rangle}\equiv  ({|{\ua_{A,i} \da _{B,i}}\rangle} + {|{ \da_{A,i} \ua _{B,i}}\rangle})/\sqrt{2}$, where ${|{\ua}\rangle}$, ${|{\da}\rangle}$ are the eigenstates of $\hat\sigma^z$ with eigenvalues $\pm1$, respectively.
The $\hat{A}_{s}$ matrices are taken to be translationally invariant (i.e., identical on every site) but the full state $|\psi_{\rm CQA}\rangle$ is not translationally invariant due to the boundary vectors $\vb{v_L}$ and $\vb{v_R}$. 
Inspired by explicit solutions found for very small system sizes, we make the ansatz that the only allowed spin states are the triplet subspace $s_i \in \{0_i,1_i,T_i\}$. Then, we use the dark state conditions Eq.~\eqref{eq-app:dark-condition} to determine the MPS coefficients of $\hat A_{s}$, $\vb{v_L}$ and $\vb{v_R}$. 

The first dark state condition $\hat{L}_{\rm AB} |\psi_{\rm CQA}\rangle = 0$ imposed by the jump term requires $(\hat{\sigma}_{A,1}^- - \hat{\sigma}_{B,1}^-) \ket{s_1} = 0$. This restricts the first site to $s_1 \in \{ 0_1 , T_1\}$. This imposes the constraint \begin{equation}
    \vb{v_L^\dagger} \hat{A}_{1} = 0. \label{eq-app:dark-jump}
\end{equation}
We are free to choose the left vector $\vb{v_L}$ and we take it to be $\vb{v_L}=\ket{0}$.

The second dark state condition is $\hat{H}^{(AB)}|\psi_{\rm CQA}\rangle = 0$. This requires that the state be a zero-energy eigenstate of the Hamiltonian $\hat H^{(AB)}$. Using our ansatz that each site $i$ of the double-system chain contains only $s_i\in\{0_i,1_i,T_i\}$ (but no $T_1$ on site 1), we can see, mechanistically, how a dark state of $\hat H^{(AB)}$ can be constructed: each action of $\hat H^{(AB)}$ on any state containing only the allowed dimer states $s_i$ always produces a single ``defect" dimer state $|S_i\rangle$. Since this state is excluded from our ansatz, $|\psi_{\rm CQA}\rangle$ must be a zero-energy eigenstate constructed so that all defects produced by the action of $\hat H^{(AB)}$ destructively interfere. As an illustrative example, consider the boundary coherent drive term $\hat H_{\rm d}^{(AB)} = \frac{\Omega}{2}(\hat\sigma^x_{A,N} - \hat\sigma^x_{B,N})$. It annihilates $|T_N\rangle$ and creates defects from $|0_N\rangle$ and $|1_N\rangle$:
\begin{equation}
    \hat{H}_{\rm d}^{(AB)} \ket{0_N} =  \frac{\Omega}{\sqrt{2}} \ket{S_N}, \quad  \hat{H}_{\rm d}^{(AB)} \ket{1_N} = - \frac{\Omega}{\sqrt{2}} \ket{S_N}, \quad \hat{H}_{\rm d}^{(AB)} \ket{T_N} = 0 .\label{eq-app:dark-drive} 
\end{equation}
One finds similar results for the dissipative boundary exchange Hamiltonian $\hat{H}_{\rm c} = -\frac{i\gamma}{2} (\hat{\sigma}_{A,1}^+ \hat{\sigma}_{B,1}^- - \hat{\sigma}_{A,1}^- \hat{\sigma}_{B,1}^+)$, and one finds that the action of the bulk Hamiltonian $\hat H_{\rm XXZ}^{(AB)} = \hat H_{\rm XXZ}^{(A)} - \hat H_{\rm XXZ}^{(B)}$ on each adjacent pair $|s_is^\prime_{i+1}\rangle$ is either zero or it generates a single defect.

The creation of these defects and the zero-energy eigenstate condition requiring that all defects destructively interfere, taken together, can be shown to constrain the form of the MPS matrices $\hat A_s$ in \emph{exactly} exactly the same way as the so-called ``isolated defect operator" method employed in Refs.~\cite{Prosen2011-uf,Prosen2015-vk}. The CQA dark state condition $\hat L^{(AB)}|\psi_{\rm CQA}\rangle=0$ is identical to the dissipative boundary condition found in Ref.~\cite{Prosen2011-uf}, but it provides a more physically transparent reason for the boundary condition. Following Ref.~\cite{Prosen2011-uf}, we choose a matrix representation $\mathcal{V}$ in the semi-infinite space $\{|k\rangle:k=0,1,\dots\}$ with the three (translationally-invariant) matrices
\begin{equation}
    \hat{A}_T = \sum_{k=0}^\infty a_k \ket{k}\bra{k}, \quad \hat{A}_1 = \sum_{k=0}^\infty b_k \ket{k+1}\bra{k},\quad \hat{A}_0 = \sum_{k=0}^\infty c_k \ket{k}\bra{k+1}, \label{eq-app:definite-A}
\end{equation} 
with coefficients $a_k$, $b_k$, and $c_k$ determined by the recursion relations coming from the Hamiltonian eigenstate condition in the bulk. Note that while these matrices are formally defined in an infinite-dimensional space, for any finite length chain $N<\infty$ we can cut the space off at level $k=N$ because the left vector $\vb{v_L}=|0\rangle$ is localized to $k=0$, hence $\bra{0}(\hat A_0)^N\ket{k>N}=0$. Working in units of the hopping rate $J\equiv1$, we obtain two recursion relations that determine $a_k$ and the product $b_k c_k$:
\begin{align}
    &a_{k+1} - 2\Delta a_{k} + a_{k-1} = 0, \label{eq-app:bulk-a-recursion}\\
    &b_kc_k - b_{k-1} c_{k-1} = a_k (\Delta a_{k} - a_{k-1}). \label{eq-app:bulk-bc-recursion}
\end{align}
The initial conditions for $a_0$, $a_1$, and $b_0 c_0$ are determined by the dissipative boundary.
The dark state condition Eq.~\eqref{eq-app:dark-jump} is automatically satisfied by picking the left vector $\vb{v_L} = \ket{0}$ and $\hat{A}_1$ a raising operator, but the exchange Hamiltonian $\hat H_{\rm c}$ generates extra defect terms in the Hamiltonian dark state condition for site 1 (compared to sites only acted upon by the bulk Hamiltonian) which, in turn, impose the initial conditions
\begin{align}
    a_0 = 1,\quad a_1 = \Delta + \frac{i\gamma}{2},\quad c_0 b_0 = \frac{i\gamma}{2}. \label{eq-app:a-bc-recursion-init-cond}
\end{align}
In the easy-axis regime $\Delta/J < 1$ and with the above initial conditions, the recursion relations for $a_k$ and the product $b_k c_k$ have the exact analytic solution
\begin{align}
    &a_k = \cos (k\eta) + \frac{i\gamma}{2} \frac{\sin (k\eta)}{\sin \eta}, \label{eq-app:a-analytical}\\
    & b_k c_k= \frac{1}{2}\sin[(k+1)\eta] \left[\frac{i\gamma}{\sin \eta} \cos (k\eta) - (1+ \frac{\gamma^2}{4\sin^2\eta})\sin(k\eta) \right],\label{eq-app:b-analytical}
\end{align}
where $\cos \eta \equiv \Delta/J$. The dissipation dark state condition $\hat L^{(AB)}|\psi_{\rm CQA}\rangle=0$ is satisfied, and the  Hamiltonian zero-energy eigenstate condition is satisfied on all sites except the coherently driven site $N$.

To fully satisfy the energy eigenstate condition, we must find the coefficients $\alpha_k$ of the right vector $\vb{v_R} = \sum_{k = 0}^N \alpha_k \ket{k}$. The interference between the defects generated by the bulk Hamiltonian $\hat H_{\rm XXZ}^{(AB)}$ and the coherent drive $\hat H_{\rm d}^{(AB)}$ yields the recursion relation
\begin{equation}
    c_k \alpha_{k+1} + \frac{\sqrt{2}}{\Omega}(\Delta a_{k} - a_{k+1}) \alpha_k - b_{k-1} \alpha_{k-1} = 0, \quad {\rm for}\; k > 0. \label{eq-app:alpha-recursion}
\end{equation}
We take the initial conditions for the recursion relation to be $\alpha_0  = 1$ and  $\alpha_{-1}=0$. Then we have a well-defined recursion relation, but one that depends on the coefficients $b_k$ and $c_k$ separately. Notice that the bulk recursion relation Eq.~\eqref{eq-app:bulk-bc-recursion} only determines the product $b_k c_k$, which reflects a gauge freedom in the MPS representation, whereas Eq.~\eqref{eq-app:alpha-recursion} requires $b_k$ and $c_k$ separately. In fact, we can always perform a similarity transformation $\hat V$ so that $\Tilde{\hat{A}}_s = \hat{V} \hat{A}_s \hat{V}^{-1}$, and $\Tilde{\vb{v}} = \hat{V} \vb{v}$. While in general, $\hat V$ can be any invertible matrix, here we consider the restricted similarity transformation $\hat{V} = \sum_{k=0}^N v_k \ket{k}\bra{k}$ . This specific transformation leaves the MPS matrices still in tridiagonal form \cref{eq-app:definite-A}. The transformation $\hat{V}$ leaves the physical state $\ket{\psi_{\rm CQA}}$ invariant. Yet it transforms the auxiliary operator and right vector coefficients to
\begin{equation}
    \Tilde{c}_k = c_k \frac{v_{k+1}}{v_k},\quad\Tilde{b}_k = b_k \frac{v_k}{v_{k+1}}, \quad\Tilde{a}_k = a_k, \quad\Tilde{\alpha}_k = \frac{1}{v_k} \alpha_k , \label{eq-app:mps-gauge}
\end{equation}
while keeping the recursion relations Eqs.~\eqref{eq-app:bulk-a-recursion}, \eqref{eq-app:bulk-bc-recursion} and \eqref{eq-app:alpha-recursion} valid. As we will see, this gauge freedom allows us to find a particularly simple solution to the recursion relation Eq.~\eqref{eq-app:alpha-recursion} in the weak dissipation limit.

\subsection{Proof of steady state uniqueness} 
In this section we briefly present the proof for the uniqueness of the NESS using Frigerio's first theorem \cite{Zhang2024-jpa,Frigerio1977-math}. The theorem states that if the master equation has a full rank steady state $\hat{\rho}_{\rm ss}$, and if the set $\{\hat L_j,\hat L_j^\dagger,\hat H\}$ of jump operators $\{\hat L_j\}$ and Hamiltonian $\hat H$ generate all operators in the Hilbert space, then the steady state is unique.

First we must compute the exact steady state density matrix $\hat\rho_{\rm ss}$ of the original master equation \eqref{eq-app:sysAqme}. It is found from $|\psi_{\rm CQA}\rangle$ by tracing out the absorber
\begin{equation}
    \hat{\rho}_{\rm ss} = \Tr_B (\ketbra{\psi_{\rm CQA}}{\psi_{\rm CQA}}) = \Tr_B (\hat{U}_B \ketbra{\psi_{\rm CQA}}{\psi_{\rm CQA}} \hat{U}_B^\dagger).
\end{equation}
Here it is convenient to perform a spin-flip of all spins in the absorber system by acting with the spin-flip unitary $\hat U_B = \prod_i \hat{\sigma}_{B,i}^x$. As we show next, this spin flip of the absorber allows us to write the steady state $\hat\rho_{\rm ss}$ as a Cholesky decomposition, $\hat \rho_{\rm ss} = \hat\Psi\hat\Psi^\dagger$ where $\hat\Psi$ is a lower triangular matrix. Using this decomposition, we then prove that the steady state is full rank.

We define the matrix $\Psi_{s_a,s_b} = \bra{s_a, s_b} \hat{U}_B\ket{\psi_{\rm CQA}}$ with matrix elements indexed by $s_a=(s_{a,1},s_{a,2},\dots,s_{a,N})$ and $s_b=(s_{b,1},s_{b,2},\dots,s_{b,N})$ where $s_{a/b,i}\in\{{\ua},{\da}\}$ are the spin states of the system A and absorber B, respectively. With the definition we find the matrix elements of $\hat\rho_{\rm ss}$ to be
\begin{align}
    \bra{s_a} \hat{\rho}_{\rm ss} \ket{s_a^\prime} = \sum_{s_b} \bra{s_a,s_b} \hat{U}_B \ket{\psi_{\rm CQA}} & \bra{\psi_{\rm CQA}} \hat{U}^\dagger_B \ket{s_a^\prime,s_b} = \sum_{s_b} \Psi_{s_a,s_b} \Psi_{s_b,s_a^\prime}^* \\
    \implies  \hat{\rho}_{\rm ss} &= \hat{\Psi}\hat{\Psi}^\dagger.
\end{align}
The matrix elements of $\hat\Psi$ can be found explicitly by substituting Eq.~\eqref{eq-app:solution} into $\Psi_{s_a,s_b} = \bra{s_a, s_b} \hat{U}_B\ket{\psi_{\rm CQA}}$ to get
\begin{equation}
    \hat{\Psi} = \sum_{s_i\in\{\pm,0\}} (\vb{v_L^\dagger} \hat{A}_{s_1} \cdots \hat{A}_{s_N} \vb{v_R}) \times \hat{\sigma}_1^{s_1} \otimes \cdots \otimes \hat{\sigma}_N^{s_N}, \label{eq-app:cholesky-psi}
\end{equation}
where the $2$ by $2$ matrices are $\hat{\sigma}^+ = \ketbra{\ua}{\da}$, $\hat{\sigma}^- = \ketbra{\da}{\ua}$, and $\hat{\sigma}^0 = \frac{1}{\sqrt{2}} (\ketbra{\ua}{\ua}+ \ketbra{\da}{\da})$. These directly correspond to the double chain dimer state $|1_i\rangle$, $|0_i\rangle$, and $|T_i\rangle$, respectively. Likewise, the relabeling of the spin indices applies to the $\hat A_s$ matrices: $s\in\{0,1,T\} \to s\in \{-,+,0\}$ [cf. Eq.~\eqref{eq-app:definite-A}].
To show that $\hat \Psi$ is a lower triangular matrix, recall that $\vb{v_L^\dagger}\hat A_+=0$ and $\vb{v_L^\dagger}\hat A_0=\vb{v_L^\dagger}$. Then, we see that Eq.~\eqref{eq-app:cholesky-psi} \emph{cannot} have terms where a $\hat\sigma_i^+$ precedes the first $\sigma_{j>i}^-$. Hence, $\Psi$ is a lower triangular matrix; therefore, it is the Cholesky decomposition of $\hat \rho_{\rm ss}$.

Equipped with the NESS $\hat \rho_{\rm ss} = \hat\Psi\hat\Psi^\dagger$, we can now prove that it is full rank (i.e., positive definite).
The matrix elements of $\hat\Psi$ are directly connected to the MPS solution via
\begin{equation}
\bra{\underline{s}^\prime} \hat\Psi \ket{\underline{s}} = \vb{v_L^\dagger} \hat{A}_{s_1^\prime - s_1} \hat{A}_{s_2^\prime - s_2} \cdots \hat{A}_{s_N^\prime - s_N} \vb{v_R}.
\end{equation}
Here $\underline{s}=(s_1,\dots,s_N)$ is a configuration vector in the spin basis, with each $s_i\in\{0,1\}$. We see by direct calculation that the diagonal elements of $\hat\Psi$ are $ \bra{\underline{\nu}} \hat\Psi \ket{\underline{\nu}} = \vb{v_L^\dagger} (\hat A_{0})^N \vb{v_R}  = a_0^N \alpha_0 = 1$. Since $\hat\Psi$ is lower triangular, its eigenvalues are all 1, hence it is full rank. Therefore, for an arbitrary state $|\phi\rangle\neq 0$, we have
\begin{align}
    \langle\phi|\hat{\rho}_{\rm ss}|\phi\rangle = \langle\phi|\hat\Psi\hat\Psi^\dagger|\phi\rangle = ||\hat\Psi^\dagger|\phi\rangle||^2 > 0,
\end{align}
where the strict inequality follows from $\hat \Psi^\dagger$ being full rank. Therefore, we have shown $\hat\rho_{\rm ss}$ to be full rank.

Next, we show that $\{\hat{H}_{\rm XXZ},\hat{\sigma}_1^-,\hat{\sigma}_1^+\}$ generates the whole algebra of the Hilbert space. Note that while the full Hamiltonian includes the coherent driving $\hat H = \hat H_{\rm XXZ} + \hat H_{\rm d}$, the driving term is redundant for this proof. First, observe that
\begin{equation}
    \hat{\sigma}_2^- = \frac{1}{4} \hat{\sigma}_1^z [\hat{\sigma}_1^- ,[\hat{H}_{\rm XXZ}, \hat{\sigma}_1^z]],
\end{equation}
where we assumed $J = 1$ for simplicity. This also generates $\hat{\sigma}_2^+$ by conjugation. In general, for all sites $j>1$, we can inductively apply
\begin{equation}
    \hat{\sigma}_{j+1}^ - = \frac{1}{4} \hat{\sigma}_j^z [\hat{\sigma}_j^- ,[\hat{H}_{\rm XXZ}, \hat{\sigma}_j^z]] - \hat{\sigma}_{j-1}^-.
\end{equation}
By conjugation this also generates $\hat{\sigma}_{j+1}^+$. Since $\sigma_j^-$, $\sigma_j^+$, and their products generate the full algebra on each site, this procedure generates the full Hilbert space algebra. 

With the two conditions of Frigerio's first theorem satisfied, we have proved that $\hat\rho_{\rm ss} = \hat\Psi\hat\Psi^\dagger / {\rm Tr}(\hat\Psi\hat\Psi^\dagger)$ is the unique steady state of Eq.~\eqref{eq-app:sysAqme}.
There is one subtlety here: in the absence of a coherent drive, $\Omega = 0$, the steady state is not full rank, but rather the completely polarized pure state $|{\da\cdots\da}\rangle$. Generically, for any $\Omega\neq0$, the result holds. Finally, we point out that this proof of uniqueness extends to our previous work of the boundary driven XX chain \cite{Lingenfelter2024-xz}.

As a final note, it has been found that the ``isolated defect operator" method yields exact solutions to other dissipative XXZ chain models in the form of Cholesky decompositions of the steady states \cite{Prosen2011-uf,Karevski2013-ju,Prosen2015-vk}. Due to the similarities between the CQA approach and the previous exact solutions, both in the form of the matrix product ansatz and in the form of the solution as a Cholesky decomposition, we suspect that these previously solved models -- and others amenable to the isolated defect operator method -- could be solved using the CQA approach. Indeed, below in \cref{sec-app:cqa-incoherent} we construct the CQA system and find the doubled-system pure state solution for the model first solved in Ref.~\cite{Prosen2011-uf}.

\subsection{Observables}

In this section, we describe how to compute the relevant observables (magnetization and current) from the MPS solution \cref{eq-app:solution}.

The main quantity of interest, the average magnetization, is defined by
\begin{align}
    \langle {m} \rangle \equiv \frac{1}{N} \sum_{j = 1}^N \Tr (\hat{\sigma}_j^z \hat{\rho}_{\rm ss}) = \frac{1}{N} \sum_{j = 1}^N \frac{\bra{\psi_{\rm CQA}} \hat{\sigma}_j^z \ket{\psi_{\rm CQA}}}{\bra{\psi_{\rm CQA}}\ket{\psi_{\rm CQA}}},
\end{align}
where $\hat\sigma_j^z$ in the final expression can be either a system A operator or an absorber B operator. Here we compute it in terms of the exact solution MPS state Eq.~\eqref{eq-app:solution}.

The local spin operators $\hat\sigma_j^z$ have matrix elements $\bra{0_j} \hat{\sigma}_j^z \ket{0_j} = -1$, $\bra{T_j} \hat{\sigma}_j^z \ket{T_j} = 0$, and $\bra{1_j} \hat{\sigma}_j^z \ket{1_j} = 1$ in the triplet spin basis. All of the other terms vanish. The total magnetization is, thus, diagonal in this basis:
\begin{equation}
   m_{\{s\}} = \sum_{j = 1}^N \bra{\{s^\prime\}} \hat{\sigma}_j^z \ket{\{s\}} = \delta_{\{s\},\{s^\prime\}} (m_+ - m_-).
\end{equation}
Here $\ket{\{s\}} \equiv \ket{s_1,s_2,\cdots,s_N}$ is the spin configuration and we denote the number of $1$s and $0$s by $m_+ = \sum_{j=1}^N |\bra{1_j}\ket{\{s\}}|^2$ and $m_- = \sum_{j=1}^N |\bra{0_j}\ket{\{s\}}|^2$, respectively. We relate these quantities to the coefficients of the MPS matrices and right vector as follows.  As $\hat{A}_{0(1)}$ always move the level in the auxiliary space by $\pm 1$, the total magnetization is mapped to the distance from 0 in the auxiliary space after taking $N$ steps (e.g., the completely polarized configuration $\{{\da\cdots\da}\}$ with total magnetization $-N$ connects $\vb{v_L^\dagger}=\langle 0|$ on the left to the $\alpha_N|N\rangle$ component of the right vector). The coefficient of configuration $\{s\}$ with $m_{\{s\}} = k$ is therefore $\vb{v_L^\dagger} \hat{A}_{s_1} \cdots \hat{A}_{s_N}|k\rangle\langle k| \vb{v_R} = \bra{0} \hat{A}_{s_1} \cdots \hat{A}_{s_N} \ket{k} \alpha_k$. The magnetization density can be given compactly as
\begin{align}
    \langle \hat{m} \rangle &=- \frac{1}{N} \frac{\sum_{k=0}^N \sum_{\{s\}} k |\bra{0} \hat{A}_{s_1} \cdots \hat{A}_{s_N} \ket{k} |^2 |\alpha_k|^2 }{\sum_{k=0}^N \sum_{\{s\}} |\bra{0} \hat{A}_{s_1} \cdots \hat{A}_{s_N} \ket{k} |^2 |\alpha_k|^2} = - \frac{1}{N} \frac{\sum_{k=0}^N  k \bra{0,0} \hat{T}^N \ket{k,k} |\alpha_k|^2 }{\sum_{k=0}^N \bra{0,0} \hat{T}^N \ket{k,k} |\alpha_k|^2}.
\end{align}
Here we defined the vectorized auxiliary space $\mathcal{V}\otimes \mathcal{V}$ by mapping $\ket{k}\bra{k}\to \bra{k,k}$, and we introduced the transfer matrix $\hat{T} = \sum_{s} (\hat{A}_{s} \otimes \hat{A}_{s}^*)$. 

Because $\hat{T}$ always keeps the state on the diagonal $\ket{j,j}$, we can further reduce the vectorized auxiliary space to its diagonal subspace, with states denoted by $\ket{\vb{j}} \equiv \ket{j,j}$. By doing so, we find the magnetization density to be
\begin{equation}
    \langle \hat{m} \rangle = - \frac{1}{N} \frac{\sum_{k=0}^N  k \bra{\vb{0}} \hat{T}^N \ket{\vb{k}} |\alpha_k|^2 }{\sum_{k=0}^N \bra{\vb{0}} \hat{T}^N \ket{\vb{k}} |\alpha_k|^2} = - \sum_{k=0}^N \frac{k}{N} p_k, \label{eq-app:ave-mag}
\end{equation}
where the transfer matrix $\hat T$ is defined by 
\begin{align}
    \hat{T} \equiv \sum_{k=0}^N |a_k|^2 \ket{\vb{k}} \bra{\vb{k}} + |b_k|^2 \ket{\vb{k+1}} \bra{\vb{k}} + |c_k|^2\ket{\vb{k}} \bra{\vb{k+1}}. \label{eq-app:transfer-matrix-def}
\end{align}
\cref{eq-app:ave-mag} expresses the magnetization as a weighted average over the probability distribution $p_k = \bra{\vb{0}} \hat{T}^N \ket{\vb{k}} |\alpha_k|^2 / (\sum_{k=0}^N \bra{\vb{0}} \hat{T}^N \ket{\vb{k}} |\alpha_k|^2)$. 
Given the exact form of $a_k$, $b_k$, and $c_k$ found from the solutions Eqs.~\eqref{eq-app:a-analytical} and \eqref{eq-app:b-analytical}, and by solving Eq.~\eqref{eq-app:alpha-recursion}, we see that the statistical weight comes from two parts: the boundary contribution $|\alpha_k|^2$, which represents the interplay between drive and bulk properties, and the transfer matrix $\bra{\vb{0}} \hat{T}^N \ket{\vb{k}}$, which reflects the bulk dynamics and the dissipation. As we show in \cref{sec-app:weak-dissipation}, in the weak dissipation limit $\gamma/J \ll 1$, there is an MPS gauge in which the distribution $\{p_k\}$ takes a relatively simple form.

Similarly, there are other quantities can be defined in the vectorized auxiliary space $\mathcal{V}\otimes \mathcal{V}$. Here we omit the detailed calculation and directly state the results. By defining the partition function $Z_N = \sum_{k=0}^N \bra{\vb{0}} \hat{T}^N \ket{\vb{k}} |\alpha_k|^2$, the spin-spin correlation is found to be $\langle \hat{\sigma}^z_a \hat{\sigma}^z_b \rangle = \frac{1}{Z_N}\sum_{k=0}^N  \bra{\vb{0}} \hat{T}^{a-1} \hat{Z} \hat{T}^{b-a} \hat{Z} \hat{T}^{N-b}  \ket{\vb{k}} |\alpha_k|^2 $
for $b>a$ and where $\hat Z = \sum_k |b_k|^2 |\vb{k+1}\rangle\langle\vb{k}| - |c_k|^2 |\vb{k}\rangle\langle\vb{k+1}|$. Similarly, the spin current is $\langle j \rangle = \frac{\gamma}{J} \frac{Z_{N-1} }{Z_N}$. We see that the complexity of calculating observable quantities is equivalent to diagonalizing the $\hat{T}$ matrix, with a total time complexity of $\mathcal{O}(N^4)$. This allows us to readily calculate steady state properties for systems with up to thousands of spins.

\section{Phase transition}

\subsection{Quantum to classical mapping}

In this section we show that the transfer matrix of Eq.~(\ref{eq-app:transfer-matrix-def}) can be interpreted as the transition matrix of a classical random walk model on a one-dimensional semi-infinite lattice. Specifically, we can interpret the matrix elements $\bra{\vb{k}}(\hat{T}^\dagger) {}^N \ket{\vb{0}}$ as being the probability that a particle initialized on site 0 ends on site $k$ after $N$ time steps (with each time step corresponding to a random hop described by $T^\dagger$). Because the coefficients $\bra{\vb{k}}(\hat{T}^\dagger) {}^N \ket{\vb{0}}$ directly help determine the average magnetization via Eq.~\eqref{eq-app:ave-mag}, we can then reinterpret the magnetization as the mean distance that a particle initialized on site 0 travels in $N$ time steps, weighted by a post-selection distribution $|\alpha_k|^2$.

To map the problem of calculating $\bra{\vb{k}}(\hat{T}^\dagger) {}^N \ket{\vb{0}}$ to a classical random walk on the lattice one-dimensional semi-infinite lattice comprising $\{\ket{\vb{k}}\!,\, \vb{k}= 0,1,\cdots \}$, we can define a classical Markov process with the master equation:
\begin{equation}
    p(k,t+1) = |a_{k}|^2 p(k,t) + |c_{k-1}|^2 p(k-1,t) + |b_{k}|^2 p(k+1,t). \label{eq-app:classical-me}
\end{equation}
Heuristically, at each discrete time step $t$ a particle on site $k$ of the auxiliary lattice can either hop to the left with with probability $|b_{k-1}|^2$, hop right with probability $|c_{k}|^2$, or remain on the same site with probability $|a_k|^2$.
The probability distribution $p(k,N)$, after taking $N$ steps from the initial condition $p(k,0) = \delta_{k,0}$, encodes the transfer matrix elements $\bra{\vb{k}}(\hat{T}^\dagger) {}^N \ket{\vb{0}} = p(k,N)$. 

For Eq.~\eqref{eq-app:classical-me} to meaningfully represent a stochastic process, we must enforce probability conservation of the coefficients [cf.~Eq.~\eqref{eq-app:definite-A}]: $|a_k|^2+|b_{k-1}|^2+|c_{k}|^2=1$. This can always be enforced using the gauge freedom of the solution \cref{eq-app:mps-gauge}. First notice from the recursion relations Eqs.~\eqref{eq-app:bulk-a-recursion} and \eqref{eq-app:bulk-bc-recursion} that we can freely scale all coefficients by a constant factor $C$:
\begin{align}
    \tilde a_k = C a_k,\quad \tilde b_k = C b_k,\quad \tilde c_k = C c_k.
\end{align}
We have the gauge freedom in Eq.~\eqref{eq-app:mps-gauge} to fix the ratio $|b_k|^2/|c_k|^2$ through a judicious choice of $\{v_k\}$. Therefore, probability conservation can be enforced by solving
\begin{align}
\label{eq:stochasticity-cond}
    C^2 \left[ |a_k|^2 + \frac{\abs{v_{k-1}}^2}{\abs{v_k}^2}|b_{k-1}|^2 + \frac{\abs{v_{k+1}}^2}{\abs{v_{k}}^2}|c_k|^2  \right] = 1 \qc \forall k=0,1,\cdots N-1.
\end{align}
Note that we only need to satisfy \cref{eq:stochasticity-cond} for $k=0,1,\cdots N-1$. This is because we are only interested in the master equation \cref{eq-app:classical-me} subject to the initial condition $p(k,0) = \delta_{k,0}$ and for times $t\le N$.

We can solve \cref{eq:stochasticity-cond} by starting from $k=0$ (note that $b_{-1}=0$) and iteratively solving every equation for the ratio $\frac{\abs{v_k}^2}{\abs{v_{k+1}}^2}$ until $k=N-1$ with
\begin{equation}
     \frac{\abs{v_{k+1}}^2}{\abs{v_{k}}^2} = (|c_k|^2)^{-1} \qty(1/C^2 - |a_k|^2 - \frac{\abs{v_{k-1}}^2}{\abs{v_k}^2}|b_{k-1}|^2),
\end{equation}
with the initial condition $|v_1|^2/|v_0|^2 = (|c_0|^2)^{-1}(1/C^2 - |a_0|^2)$. It is important of course that the solutions satisfy $\frac{\abs{v_k}^2}{\abs{v_{k+1}}^2} >0$. This can guaranteed by properly choosing $C$. We start from a specific gauge $c_k = 1$. For bounded sequences $|a_k|^2 < a^2$ and $|b_k|^2 < b^2$, we pick a constant $C$ to be
\begin{equation}
      1/C^2 = 1+ a^2 +b^2.
\end{equation}
Such choice of $C$ results in $|v_1|^2/|v_0|^2 = (1/C^2 - |a_0|^2) > (1+b^2) >1$. Next we show that $|v_{k+1}|^2/|v_k|^2 >1 $ using induction. Given $|v_{k}|^2/|v_{k-1}|^2 >1 $, we have
\begin{equation}
    \frac{\abs{v_{k+1}}^2}{\abs{v_{k}}^2} =  \qty(1/C^2 - |a_k|^2 - \frac{\abs{v_{k-1}}^2}{\abs{v_k}^2}|b_{k-1}|^2) > \qty(1 + (1- \frac{\abs{v_{k-1}}^2}{\abs{v_k}^2}) b^2 )> 1,
\end{equation}
By doing so, we show $|v_{k+1}|^2/|v_k|^2 >1 $ by the specific construction of $C$. The net result is that we can always pick a gauge such that the relevant elements of the $T^\dagger$ matrix correspond to a valid transition matrix of a classical 1D hopping model.  While the procedure we have described is applicable for all parameter choices of our model, there is no guarantee that this can be done in a manner that yields closed-form analytic solutions.  
In the next section, we 
show that in the special case of weak dissipation ($\gamma \rightarrow 0$), this mapping to a classical model can be done analytically.

\subsection{Weak dissipation limit}\label{sec-app:weak-dissipation}

In this section we analyze the exact solution in the weak dissipation limit $\gamma/J\ll 1$ to obtain analytic insight into the phase transition in $\Omega$. Recall that we have exact analytic expressions for the MPS coefficients $a_k$ and $b_kc_k$ in the $\Delta/J<1$ regime given by Eqs.~\eqref{eq-app:a-analytical} and \eqref{eq-app:b-analytical}, the recursion relation Eq.~\eqref{eq-app:alpha-recursion} for $\alpha_k$, and a gauge freedom in choosing $b_k$ and $c_k$. In the weak dissipation limit, a convenient gauge is
\begin{align}
    & b_k =\frac{1}{\sqrt 2}\sin[(k+1) \eta], \label{eq-app:gauge-b} \\
    & c_k=\frac{1}{\sqrt 2}\left[\frac{i\gamma}{J\sin \eta} \cos (k\eta) - (1+ \frac{\gamma^2}{4J^2\sin^2\eta})\sin(k\eta) \right], \label{eq-app:gauge-c}
\end{align}
where $\cos\eta = \Delta/J$ and we have restored $J$. Using these solutions and Eq.~\eqref{eq-app:a-analytical} for $a_k$, and keeping only the $\gamma$-independent terms, the recursion relation Eq.~\eqref{eq-app:alpha-recursion} simplifies to
\begin{equation}
   \alpha_{k+1} - \frac{2 J \sin \eta}{\Omega} \alpha_k + \alpha_{k-1} = 0 \label{eq-app:Chebyshev-alpha}
\end{equation}
with initial conditions $\alpha_0 = 1$ and $\alpha_1 = \sin \eta /\Omega$. Here we assumed $\sin k\eta \neq 0$. This is generically true except at the set of special points discussed in Sec.~\ref{sec-app:special-points}. 

The solution to Eq.~\eqref{eq-app:Chebyshev-alpha} is a Chebyshev polynomial of the first kind $\alpha_k = T_k(\frac{J}{\Omega}\sin\eta)$. Across the critical driving strength $\Omega_c\equiv J\sin\eta = \sqrt{J^2-\Delta^2}$, there is a change in the functional form of the solution:
\begin{equation}
    |\alpha_k|^2 = \begin{cases}
        \cos^2 k\theta,~~\, \theta = \arccos (\Omega_c/\Omega) & \frac{\Omega}{\Omega_c} >1,\\
        \cosh^2 k\theta,~ \theta = {\rm arccosh} (\Omega_c/\Omega) & \frac{\Omega}{\Omega_c} <1.
    \end{cases} \label{eq-app:weak-alpha}
\end{equation}
Above the critical drive $\Omega/\Omega_c>1$, we have $|\alpha_k|^2 \leq 1$ and oscillatory behaviour as a function of $k$. Below the critical drive $\Omega/\Omega_c < 1$, the $|\alpha_k|^2$ grow exponentially in $k$. This transition in the behavior of $|\alpha_k|^2$ leads to non-analytic behavior in the spin chain magnetization and the free energy of the classical model later shown in \cref{sec-app:free-energy}.

\begin{figure}
    \centering
    \includegraphics{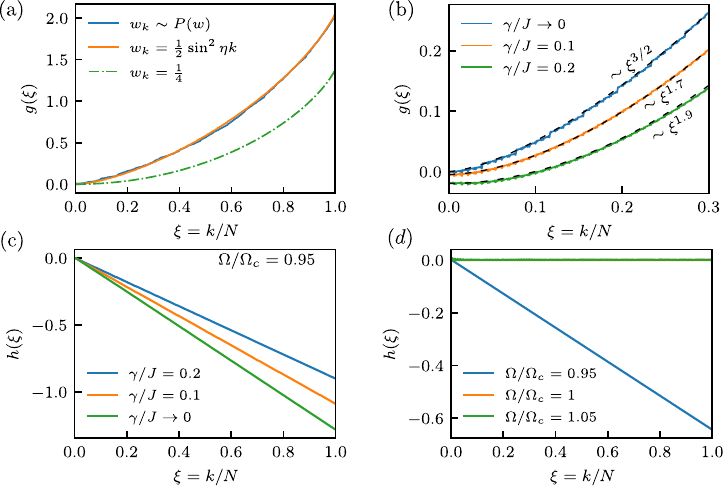}
    \caption{(a) $g(\xi)$, the 
    log of the $N$-step transfer matrix element from site $0$ to $k$, c.f.~Eq.~(\ref{eq:g-def}), is plotted versus $\xi = k / N$.  This function plays the role of a potential in our effective free energy.   The quasi-periodic distribution coming from the exact solution (orange line) is well approximated by the random environment with the same weight distribution $P(w)$ (blue line). The results from a simplified model with uniform hopping $w_k = 1/4$ (green dashed line) yields diffusive behaviour, and does not agree with the full model.  (b) The scaling of the potential $g(\xi)$ with $\xi$ is shown near the origin. For $\gamma \rightarrow 0$, it follows  $\xi^{3/2}$ (dashed line).  As the dissipation $\gamma$ is increased, we find that the power law changes continuously.  (c) Comparing field term $h(\xi)$ for different dissipation. Finite dissipation leads to different slope. (d) Field term $h(\xi)$ in the free energy $F(\xi) = g(\xi) + h(\xi)$ are plotted vs. $\xi$ using the exact solution ($\gamma/J = 10^{-3}$). Here $h(\xi)$ changes across the critical drive from $0$ above $\Omega_c$ to a linear field below.}
    \label{fig-app:transfer-matrix}
\end{figure}

The transfer matrix has a particularly simple form in the weak dissipation limit; this allows us to derive insights about the phase transition from an effective classical random walk model. Keeping only the zeroth order in $\gamma$ terms in $a_k$, $b_k$, and $c_k$ [except for $c_0$ whose leading order is $\mathcal{O}(\gamma^2)$, see discussion at the end of Sec.~\ref{sec-app:weak-dissipation}], we see that the classical master equation \eqref{eq-app:classical-me} 
in this limit yields simple reciprocal hopping rates that vary quasiperiodically along the auxiliary lattice:
$|b_{k-1}|^2=|c_k|^2 = w_k \equiv \frac{1}{2}\sin^2 k\eta$ and $|a_k|^2 = 1-2 w_k$. Hence the classical master equation is
\begin{equation}
    p(k,t+1) = (1-2w_k) p(k,t) + w_{k-1} p(k-1,t) + w_{k+1} p(k+1,t).
\end{equation}
In general the effective wavelength $2 \pi / \eta$ of the modulation of the $w_k$ will not be commensurate with the auxiliary lattice, so the classical master equation corresponds to a model of quenched disorder.  
We focus on the final distribution after $N$ time steps, $f_N(k) \equiv p(k,N) = \bra{\vb{k}} (\hat{T}^\dagger)^N \ket{\vb{0}}$.
To get insight into the physics of our classical model, consider first
a related, simplified model where we ignore disorder, and assume $w_k = \overline{w_k} = \frac{1}{4}$. We find for this case $f_N(k) = \frac{1}{2^N}\exp(-N g(k/N))$, where $g(\xi)=2\xi\,{\rm artanh}(\xi)+\ln(1-\xi^2)$ for the reduced coordinate $\xi = k/N$. As we see in Fig.~\ref{fig-app:transfer-matrix}(a), this result does not agree with the numerically-computed distribution using the  $w_k=\frac{1}{2}\sin^2 k\eta$ of the exact classical model. Thus, the disorder-free approximation is not sufficient to obtain the correct functional form of $f_N(k)$.

Noting that the weights $w_k = \frac{1}{2}\sin^2 k\eta$ are quasi-periodic in $k$, we expect that for $\eta\neq \pi \frac{l}{m}$, the arguments of the sin function, $(k\eta\,{\rm mod}\,\pi)$ are distributed uniformly in the interval $[0,\pi)$. Thus, we make the approximation that the weights are distributed in the interval $[0,1/2)$ with probability distribution
\begin{align}
    P(w) = \frac{2}{\pi\sqrt{2w(1-2w)}}.
\end{align}This classical single-particle model is similar to the Sinai random walk \cite{Sinai-1983-randomwalk}, but with a crucial difference: there is a non-zero probability $1-2w_k$ to remain on site $k$.
As we see in Fig.~\ref{fig-app:transfer-matrix}(a), this model estimates the true transfer matrix elements $f_N(\xi)$ extremely well. We seek for a specific form taking out the system size contribution
\begin{equation}
\label{eq:g-def}
    g(\xi) \equiv -\frac{1}{N} \ln \bra{\vb{k}} (\hat{T}^\dagger)^N \ket{\vb{0}}.
\end{equation}
We numerically find sub-diffusive scaling $\langle k^2\rangle\sim N^{2/3}$ and $g(\xi)\sim \xi^{3/2}$, as shown in Fig.~\ref{fig-app:transfer-matrix}(b). In contrast, if we took the weights to be uniform, $w_k=\frac{1}{4}$, then we would find a diffusive behaviour, i.e.~$g(\xi)\sim\xi^2$.

Finally we turn to the behavior of the transfer matrix elements $\bra{\vb{0}} \hat{T}^N \ket{\vb{k}} =\bra{\vb{k}} (\hat{T}^\dagger)^N \ket{\vb{0}} $ [cf.~Eq.~\eqref{eq-app:transfer-matrix-def}] in thermodynamic limit $N\to\infty$. These are the matrix elements relevant for analytically describing the phase transition in the coherent drive $\Omega$.
In the weak dissipation limit, we keep only zeroth order in $\gamma$ terms for all coefficients except $c_0$ which would vanish to $\mathcal{O}(\gamma^2)$. The transfer matrix is 
\begin{equation}
    \hat{T}^\dagger = \sum_{k=0}^\infty  \cos^2 k \eta \ketbra{\vb{k}}{\vb{k}} + \frac{1}{2} \sin^2 \eta (k+1) \ketbra{\vb{k}}{\vb{k+1}} + \frac{1}{2} \sin^2 \eta k\ketbra{\vb{k+1}}{\vb{k}}  + \frac{\gamma^2}{2\sin^2 \eta} \ketbra{\vb{1}}{\vb{0}} \label{eq-app:T-mat}
\end{equation}
We see that $\bra{\vb{0}} (\hat{T}^\dagger)^N \ket{\vb{0}}=\mathcal{O}(1)$ in $\gamma$ but all other matrix elements are $\bra{\vb{k}} (\hat{T}^\dagger)^N \ket{\vb{0}}=\mathcal{O}(\gamma^2)$. 
When taking the thermodynamic and weak dissipation limits, it is important to take $N\to\infty$ while keeping $\gamma>0$. Taking this order of limits, each $k>0$ element of $\bra{\vb{k}} (\hat{T}^\dagger)^N \ket{\vb{0}}$ is at least $\mathcal{O}(N)$ whereas the $k=0$ element is $\mathcal{O}(1)$. Therefore, in the thermodynamic limit we can safely ignore the single contribution $\langle \vb{0}|(\hat T^\dagger)^N|\vb{0}\rangle$. Then we can take $\gamma\to0$, keeping leading $\mathcal{O}(\gamma^2)$ terms only.

\subsection{Free Energy and Critical Exponent}
\label{sec-app:free-energy}
In this subsection, we explain how to understand the phase transition from an effective free energy and derive the critical exponent in the weak dissipation limit. 
We can rewrite the average magnetization Eq.~\eqref{eq-app:ave-mag} in terms of a dimensionless ``free energy" $F(\xi)$
\begin{equation}
    \langle \hat{m} \rangle = -\frac{\sum_{\xi} \xi e^{-N F(\xi)} }{\sum_\xi e^{-N F(\xi)}},
\end{equation}
where $\xi=k/N\in[0,1]$ is the reduced coordinate. The free energy $F(\xi) = g(\xi) + h(\xi)$ decomposes into the $\Omega$-independent ``potential'' term $g(\xi)$ [\cref{eq:g-def}] and the $\Omega$-dependent ``external field" term 
\begin{align}
    h(\xi) \equiv - \frac{1}{N}\ln |\alpha(\xi)|^2.\label{eq-app:h-def}
\end{align}

In the thermodynamic limit, the mean magnetization is found by minimizing the free energy $F(\xi)$ over $\xi$. Using the weak dissipation expression  \cref{eq-app:weak-alpha} for $|\alpha(\xi)|^2$, we find that the field term is continuous but non-analytical in the thermodynamic limit due to the discontinuous derivative $\partial_\Omega h(\xi)$ at $\Omega_c$:
\begin{equation}
    \lim_{N \to \infty} h(\xi) = -\lim_{N \to \infty} \frac{1}{N} \ln |\alpha_\xi|^2 = \begin{cases}
        -\lim_{N \to \infty} \frac{1}{N} \ln \cos^2 N\theta \xi = 0, & \Omega>\Omega_c, \\
        -\lim_{N \to \infty} \frac{1}{N} \ln \cosh^2 N\theta \xi = -2 \theta \xi , & \Omega<\Omega_c.
    \end{cases} \label{eq-app:h-non-analytic}
\end{equation}
The non-analytical behavior of the field $h(\xi)$ explains the mechanism behind the phase transition. In the ordered phase $\Omega> \Omega_c$, the field term is turned off $h(\xi) = 0$. The free energy $F(\xi) = g(\xi)$ has a global minimum at $\xi^* = 0$. Whereas when $\Omega < \Omega_c$, the free energy takes the form of $F(\xi) = g(\xi) - 2 \theta \xi$, with a global minimum $\xi^*>0$. 

Note that in our chosen gauge, the mapping to a classical stochastic model is only valid for $\gamma \rightarrow 0$.  We stress however that this is just a question of interpretation.  The expressions for $\langle m \rangle$ and our effective free energy remain valid no matter the value of $\gamma$.  
The behavior of the external field $h(\xi)$ is numerically computed for $\gamma/J = 10^{-2}$, using the exact solution in Fig.~\ref{fig-app:transfer-matrix}(d) for $N=1000$. We see that $h(\xi)$ suddenly switches on for $\Omega < \Omega_c$.

We can obtain a critical exponent $\beta$ for the scaling of magnetization with the drive below the critical point, $\langle\hat m\rangle \sim \delta^\beta$, where $\delta = \frac{\Omega_c - \Omega}{\Omega_c}\ll1$. Near the transition, we assume the potential is a power law $g(\xi) = g_0 +g_1\xi^m$ for some $m>0$. The global minimum of the free energy is, thus,
\begin{align}
    \delta F(\xi^*) = 0 \implies \xi^* = \xi_0 \theta^{\frac{1}{(m-1)}}.
\end{align}
Recall that $\cos \theta = \Omega_c/\Omega$ for $\Omega> \Omega_c$, and $\cosh \theta = \Omega_c/\Omega$ for $\Omega< \Omega_c$.
Further, as discussed in \cref{sec-app:weak-dissipation}, we numerically found that 
$g(\xi)\sim \xi^{3/2}$ in the weak dissipation limit, implying that $m=3/2$. Therefore, we predict a critical exponent of $\beta = 1$ in the $\gamma/J\to0$ limit. 

Away from the weak dissipation limit, we numerically find that $h(\xi)$ undergoes a similar transition as in \cref{eq-app:h-non-analytic}. However, both $h(\xi)$ and $g(\xi)$ show continues dependence on $\gamma$ for $\Omega < \Omega_c$ [see \cref{fig-app:transfer-matrix}(b) and \cref{fig-app:transfer-matrix}(c)].
As we will see in the next section, the phase transition still occurs at the same point even away from weak dissipation limit, but the critical exponent $\beta$ varies continuously with $\Omega$.

\subsection{Finite size scaling}\label{sec-app:finite-size}

\begin{figure}[ht]
    \centering
    \includegraphics{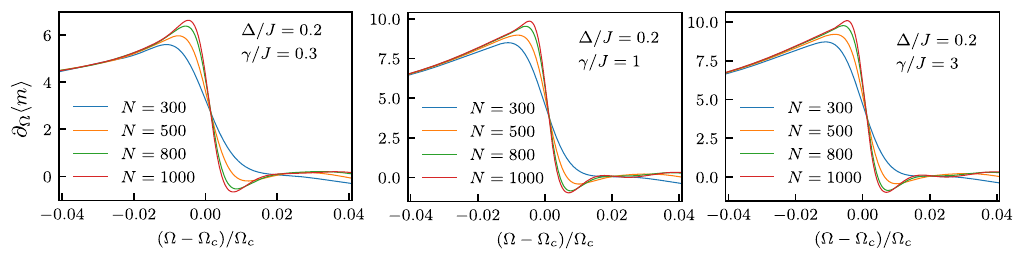}
    \caption{The magnetic susceptibility $\partial_\Omega\langle\hat m\rangle$ is plotted as a function of $\Omega$ near the critical drive $\Omega_c$. Here we use the analytic prediction for $\Omega_c = \sqrt{J^2 - \Delta^2}$. Each plot is for a different (labeled) value of dissipation $\gamma/J$. We find a small relative error on the order $10^{-3}$ between the analytically predicted $\Omega_c$ and the crossing point of the numerically calculated finite-size susceptibilities. We suspect this error could itself be a finite-size effect that slowly vanishes in the thermodynamic limit.}
    \label{fig-app:susceptibility}
\end{figure}

In this section, we perform numerical analysis of the phase transition away from the weak dissipation limit.
First, we show that the critical drive $\Omega_c = \sqrt{J^2-\Delta^2}$, derived analytically in the weak dissipation limit, remains valid even away from that limit. 
Numerically, we diagnose the critical drive by  the discontinuity in the magnetization susceptibility $\partial_\Omega \langle \hat m \rangle$ (in the weak dissipation limit, this corresponds to the sharp change in the free energy external field $h(\xi)$ in Eq.~\eqref{eq-app:h-non-analytic}). Numerically, we observe in Fig.~\ref{fig-app:susceptibility} that the susceptibility for different system sizes cross each other all at one point. Moreover, this crossing agrees extremely well with the analytically predicted $\Omega_c = \sqrt{J^2-\Delta^2}$, even a way from the weak dissipation limit.

\begin{figure}[ht]
    \centering
    \includegraphics{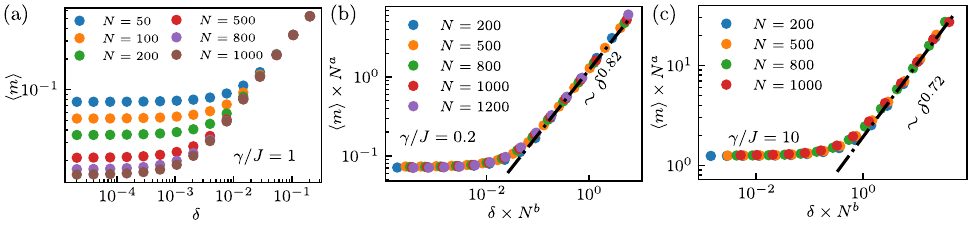}
    \caption{(a) The finite size scaling plot in the main text, Fig.~2(c), without the scaling with system size with $\delta = (\Omega_c - \Omega)/\Omega_c$. (b) and (c) Finite size scaling for weak dissipation $\gamma/J = 0.2$ and strong dissipation $\gamma/J = 10$. For panel (b), although the dissipation is weak, it is still sufficiently strong to avoid the condensation physics.}
    \label{fig-app:finitesize}
\end{figure}

Next, we perform a finite size scaling analysis and observe scaling collapse near the critical drive $\delta = \frac{\Omega_c- \Omega}{\Omega_c} \ll 1$ to a single universal scaling function:
\begin{equation}
    \langle m\rangle = N^{-a} \mathcal{M}( \delta N^b)
\end{equation}
for scaling exponents $a$ and $b$. In particular, near the transition, we find the scaling function $\langle \hat m \rangle \sim \delta^\beta N^{\beta b- a}$ where the fitted scaling exponents $a$ and $b$ are related to the observed critical exponent $\beta = \frac{a}{b}$. In Fig.~\ref{fig-app:finitesize}(a) we reproduce Fig.~2(c) of the main text \emph{without} rescaling, which verifies that the scaling collapse to the above universal function is a real effect. 
Furthermore, we see that there is clear scaling collapse to a single scaling function even as the dissipation $\gamma/J$ is varied. Despite the $\gamma$-dependence of the critical exponent $\beta$, we still find universal scaling collapse, as shown in Fig.~\ref{fig-app:finitesize}(b) and (c), and that $\beta = \frac{a}{b}$ holds.

\begin{figure}[h]
    \centering
    \includegraphics{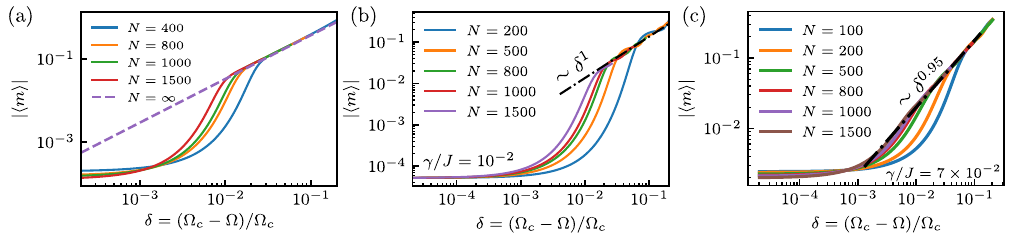}
    \caption{(a) Critical behavior for weak dissipation using the approximate expression \cref{eq-app:condensation} for $\langle \hat m\rangle$ with $\gamma/J = 10^{-2}$. We use the analytical expression for free energy and weak dissipation limit $g(\xi)\sim \xi^{3/2}$. The dashed line is the infinite size limit. This suggests we should fit to the power law scaling part to overcome the finite size effect. (b) and (c) Full numerical results of critical behavior for weak dissipation using the exact solution. Comparing with the approximation in panel (a), we see that the approximation captures the essential effects of the condensation physics for sufficiently small $\gamma/J$. By fitting to the power law part, we obtain the critical exponent $\beta$. As $\gamma/J$ decreases, the exponent approaches the analytic prediction $\beta=1$.}
    \label{fig-app:condensation}
\end{figure}

In Fig.~\ref{fig-app:finitesize} we focused on dissipation rates $\gamma/J \sim 1 $ instead of in the weak dissipation limit. There is a subtlety in analyzing the weak dissipation limit related to the weak bond in the transfer matrix $\bra{\vb{1}} \hat{T}^\dagger \ket{\vb{0}} = \frac{\gamma^2}{\sin^2 \eta}$ discussed in \cref{eq-app:T-mat}. We argued that in the thermodynamic limit, taking care order the limits $N\to\infty$ before $\gamma\to 0$, this weak bond does not matter as the contribution to the magnetization of $\bra{\vb{0}} (\hat{T}^\dagger)^N \ket{\vb{0}}/ \sum_k^N \bra{\vb{k}} (\hat{T}^\dagger)^N \ket{\vb{0}}\sim\mathcal{O}(N^{-1})$ vanishes. For finite $N$ and sufficiently weak $\gamma/J\ll 1$, this assertion is no longer valid. In fact, due to the $\mathcal{O}(\gamma^2)$ suppression of $\bra{\vb{1}} \hat{T}^\dagger \ket{\vb{0}}$ relative to $\bra{\vb{0}} \hat{T}^\dagger \ket{\vb{0}}$, we observe a kind of ``condensation" physics, wherein there is an unusually large occupation of the fully depolarized $\langle \hat m\rangle = 0$ state. In this regime, we approximate $\bra{\vb{0}}(\hat T^\dagger)^N\ket{\vb{0}} = 1+\mathcal{O}(\gamma^2) \approx 1$. Thus, the magnetization is
\begin{equation}
    \langle \hat m \rangle = - \frac{1}{N}\frac{\sum_{k>0}^N k \bra{\vb{k}} (\hat{T}^\dagger)^N \ket{\vb{0}}|\alpha_k|^2}{1 + \sum_{k>0}^N \bra{\vb{k}} (\hat{T}^\dagger)^N \ket{\vb{0}} |\alpha_k|^2} = - \frac{\gamma^2 \int \dd \xi \exp(- N F(\xi))\xi}{1+\gamma^2 \int \dd \xi \exp(- N F(\xi))}, \label{eq-app:condensation}
\end{equation}
where the condensation effects appear as the $\mathcal{O}(1)$ term in the normalization. This additional term is significant when $\gamma^2 \int \dd \xi \exp(- N F(\xi)) \approx \gamma^2 N^{2/3} \ll 1$ and vanishes by taking the thermodynamic limit first. However, we see that for small $\gamma$, we require very large system sizes to eliminate this effect. 
In Fig.~\ref{fig-app:condensation} we see strong finite-size effects for $\delta \ll 1$ due to this behavior. Fig.~\ref{fig-app:condensation}(a) is a plot of the approximate magnetization Eq.~\eqref{eq-app:condensation} for various system sizes and Fig.~\ref{fig-app:condensation}(b) and (c) show the magnetization calculated numerically from the exact solution. Nevertheless, our approximation captures the essential physics and we see the finite size effect vanishes for large system size in \cref{fig-app:condensation}(c). Moreover, this allows us to extract the critical exponent $\beta$ by fitting the scaling of $\langle \hat m\rangle\sim \delta^\beta$ in the range of $\delta$ sufficiently large to avoid the effects of the condensation physics [see Fig.~\ref{fig-app:condensation}(b) and (c)]. We confirm that as $\gamma/J\to 0$, we recover the predicted critical exponent $\beta = 1$.

\section{Special points of anisotropy \textDelta/\textit{J}}
\label{sec-app:special-points}

\subsection{No phase transition at special points}

In this section, we discuss properties of the exact solution when the anisotropy (or interaction) parameter $\Delta$ is tuned to a special point $\Delta_{l,m}$ defined by 
$\Delta_{l,m}/J = \cos \frac{l}{m} \pi \equiv \cos\eta$ for $m,l \in \mathbb{Z}$. We 
find that at these special values of $\Delta$, there is no critical behaviour as a function of $\Omega$ in the vicinity of $\Omega_c$. We can 
establish this lack of a phase transition rigorously in the limit of weak dissipation $\gamma \to 0$. Numerically, we find that the lack of critical behavior persists even for $\gamma\sim J$. 

First, using the exact solution constraint equation \cref{eq-app:b-analytical}, we immediately see that $\Delta = \Delta_{l,m}$ are special because the product of the MPS coefficients $b_{nm-1} c_{nm-1} = 0$ for every positive integer $n\in\mathbb{N}$. In general, we can always choose an MPS gauge for which $b_{nm-1} = 0$; however, for finite $\gamma>0$ we must have all $c_{k} \neq0$ in order to have a boundary state with $\alpha_{k+1}<\infty$ [cf.~Eq.~\eqref{eq-app:alpha-recursion}]. For the classical random walk model, the vanishing $b_{nm-1} = 0$ coefficients imply that a particle on site $nm$ can never hop further to the left, i.e.~to sites $k$ with $k \leq nm-1$. 

In the weak dissipation limit, a suitable gauge for which $b_{nm-1}=0$ is the same gauge given in Eqs.~\eqref{eq-app:gauge-b} and \eqref{eq-app:gauge-c}. Notice that in this gauge, $c_{nm} \propto \gamma$ and so becomes arbitrarily small for $\gamma \to 0$.  In the zero dissipation limit, we thus have that each site $nm$ in the auxiliary lattice ($n\in\mathbb{N}$) becomes an absorbing site:  there is no hopping out of this site, either to the right or the left.  Recall that our exact solution MPS corresponds to a particle starting on site $0$ of the auxiliary lattice and attempting to hop to other sites.  For $\gamma \to 0$ and for $\Delta = \Delta_{l,m}$, we thus have that a particle that starts on site 0 can never move 
past site $m$, and if it reaches this site, will become stuck there.    
Therefore, at a special point $\eta = \frac{l}{m}\pi$ (and for weak $\gamma$), the spin chain has an effective chain length $N_* = m$, even in the thermodynamic limit. This implies that there can be no phase transition at the special values of anistropy $\Delta_{l,m}$ in the zero dissipation limit.

\begin{figure}[h]
    \centering
    \includegraphics{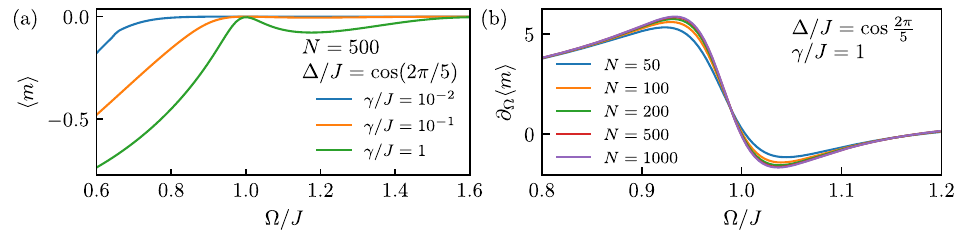}
      \caption{Special points of anisotropy $\Delta/J = \cos \frac{l}{m}\pi$ do not show critical behavior. (a) Magnetization is plotted vs. drive strength $\Omega/J$ for the special point $\eta=\frac{2}{5}\pi$ for various dissipation rates $\gamma/J$. No sharp transition in the magnetization is observed, even for a $N=500$ site chain. (b)
      For the same $\eta$ as (a), the magnetization susceptibility is plotted vs drive strength for fixed $\gamma/J=1$ but for various system sizes. Contrasting with the non-special anisotropy point $\Delta/J = 0.2$ [cf.~Fig.~\ref{fig-app:susceptibility}], here we see no evidence for a sharpening of the change in susceptibility around $\Omega_c \approx 0.95$.}
    \label{fig-app:fractal}
\end{figure}

Away from the weak dissipation limit, only the cutoff $b_{nm-1}=0$ remains.  This implies that starting in site $nm$, hopping to the left is still impossible, but there is an amplitude for hopping to the right.  Despite there no longer being a strict localization of the dynamics in the auxiliary lattice, we nevertheless have numerical evidence that there still is no critical behavior near the critical drive $\Omega_c$ at the special points even for $\gamma\sim J$. 
As we see in Fig.~\ref{fig-app:fractal}(a) and (b), there is not a sharp transition in the magnetization at $\Omega_c$. In Fig.~\ref{fig-app:fractal}(a), we see that the magnetization smoothly decreases and is highly $\gamma$-dependent for the special point $\eta = \frac{2}{5}\pi$. In Fig.~\ref{fig-app:fractal}(b), we see that for fixed $\gamma$ and $\eta = \frac{2}{5}\pi$, the magnetization susceptibility does not continue to sharpen with increasing system size. This is consistent with the notion of a effective chain length $N_* < N$. We also find in Fig.~4(a) of the main text that the entanglement entropy shows an area law at the special point $\eta=\frac{\pi}{3}$ vs. logarithmic scaling $S_{N/2}\sim \log N$ away from special points.

As a final remark, we also note that at a special point for any $\gamma$, the matrix elements $|a_k|^2 = |a_{k+m}|^2$, $|b_k|^2 = |b_{k+m}|^2$, and $|c_k|^2 = |c_{k+m}|^2$ becomes periodic with period $m$ in the auxiliary space $\mathcal{V}$. Recall that there is always a gauge in which $b_{nm-1}=0$. Together with periodicity, this reduces the transfer matrix $\hat{T}$ to an block upper triangular form:
\begin{equation}
    T = \begin{pmatrix}
        T_m & O_m & 0 & \cdots \\
        0 & T_m & O_m& \\
        0 & 0 & T_m & \\
        \vdots & && \ddots & O_m \\
        &&&& T_m
    \end{pmatrix}.\label{eq-app:fractal-tmat}
\end{equation}
Here $T^\dagger_m $ and $(O_m)_{jk} = \delta_{j,m} \delta_{k,1} |c_{m-1}|^2$ are $m\times m$ matrices. 
This transfer matrix structure leads to finite number of (degenerate) eigenvalues and exponentially decaying spatial correlations. 
We leave determining if and how exactly this structure prevents critical behavior for future investigation.

\subsection{Resonances in magnetization at special points}

In this section, we explain briefly how the resonance features, shown in Fig.~3 of the main text, emerge from the classical random walk model. For $\gamma/J \ll 1$, the transfer matrix
in the convenient gauge defined by Eqs.~(\ref{eq-app:gauge-b}) and (\ref{eq-app:gauge-c}), 
for the $p$th unit cell $\hat{T}^\dagger_{m;p} = \sum_{k=pm}^{pm+m-1}  \cos^2 k \eta \ket{\vb{k}} \bra{\vb{k}} + \frac{1}{2} \sin^2 [(k+1)\eta] \ket{\vb{k}} \bra{\vb{k+1}} + \frac{1}{2} \sin^2 k\eta \ket{\vb{k+1}} \bra{\vb{k}}  + \frac{\gamma^2}{2\sin^2 \eta} \ket{\vb{1}}\bra{\vb{0}}$. Here $T_{m;p}$ is the $p$th diagonal block $T_m$ of Eq.~\eqref{eq-app:fractal-tmat}.
To this order in $\gamma$, we have no amplitude for hopping from site $k= (pm-1 ) \to k=pm$ (though the amplitude for the reverse process is non-zero).  
To next order in $\gamma$, rightwards hopping to this bond is possible but with a very weak amplitude $\sim \gamma^2/(\sin^2\eta)$.  
For an auxiliary-space particle initialized on site 0 to reach a final coordinate $k = nm +s$, the particle has to pass a weak bond $\gamma^2/(\sin^2\eta)$ at least $n$ times. The transfer matrix is therefore $f_N(k)= \bra{\vb{k}} (\hat{T}^\dagger)^N \ket{\vb{0}} \approx C(n,N) f(s) (\frac{\gamma^2}{\sin^2\eta})^n$ to  leading order in $\gamma$. 
Here $C(n,N)$ counts the number configurations that cross $n$ weak bonds to reach $k$, and $f(s)$ is a function that depends only on the position inside of the unit cell. 
This leads to a potential that gets deeper as $\gamma\to0$. Therefore, to move away from the $\langle \hat m\rangle =0$ configuration, a larger field $\theta$ is required (thus, a correspondingly \emph{weaker} drive $\Omega$, cf.~Eq.~\eqref{eq-app:weak-alpha}). For $\gamma\ll \Omega$, we expect the magnetization to be close to zero, and we see this numerically: decreasing the dissipation results in higher resonance peaks around the special anisotropy points which agrees with our crude argument.

\subsection{Survival of the phase transition in the thermodynamic limit}

Lastly, we argue that although there are infinitely many special points $\eta = \frac{l}{m}\pi$ in the thermodynamic limit (i.e., all rationals $l/m$), the phase transition for generic points persists.  The crucial observation is that the resonances get sharper with increasing $N$. We assume that this sharpening persists so that they become arbitrarily sharp in the thermodynamic limit. At any fixed $N$, even though any generic non-special $\eta$ will fall within the resonance width of some special point, only two things can be: Either that special point will have a transfer matrix cutoff dimension $m \gtrsim N$ (in the weak dissipation limit), in which case there is no resonance. Or there is a resonance, in which case we can always increase $N$ until the generic point no longer falls within the resonance width of the problem special point.

We can be a little bit more formal, again only assuming the sharpening of resonance peaks with $N$.
Similar to the argument for the fractal points of of the Drude weight \cite{Agrawal2020-prb}, we assume the generic point $\Delta/J = \cos \lambda \pi$ where $\lambda$ is a irrational number. By Dirichlet's approximation theorem, we can always find a rational $\lambda_q = \frac{p}{q}$ that approximates $\lambda$ with an error bounded by $|\lambda_q-\lambda|\lesssim 1/q^2$. Thus, better approximations of $\lambda$ necessarily require larger $q$ for generic $\lambda$, and so have larger effective chain lengths $N_*\sim q$.
One may worry about cases where $\lambda$ is very close to a special point with a very small $q$ (e.g., $\eta = \frac{\pi}{3}$). We can always find a rational number $\lambda_s$ satisfying $\lambda < \lambda_s < \lambda_q$ (taking $\lambda<\lambda_q$) that better approximates $\lambda$, but with a greater denominator $s$. So for sufficiently long chains, any non-special anisotropy point will exhibit critical behavior around $\Omega_c$.

\section{Transport}\label{sec-app:transport}

\begin{figure}[h]
    \centering
    \includegraphics{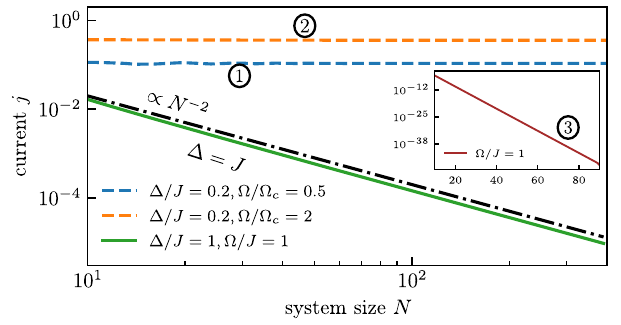}
    \caption{Transport properties of the coherent boundary driven spin chain in the three different phases [cf.~Fig.~2 of the main text]. The current $\langle j \rangle$ is plotted as a function of system size $N$. In the easy-axis regime $\Delta/J < 1$ the current is ballistic, and differs only by a prefactor between the low density phase $(1)$ and the high density phase $(2)$. At the Heisenberg point $\Delta/J=1$, the current scaling is sub-diffusive, $\langle j \rangle\sim N^{-2}$, as shown by the green curve closely following the black dashed line $\sim N^{-2}$. In the inset we plot the current in the $\Delta/J>1$ regime with $\Delta/J = 1.2$. The current falls exponentially with $N$, indicative of an insulating phase.
    For all curves we pick $\gamma/J = 1$.}
        \label{fig-app:transport}
\end{figure}

In this section, we briefly introduce the transport property of the coherent boundary driven-dissipative spin chain \cref{eq-app:sysAqme}. The spin current is defined by $\langle j \rangle \equiv {i \langle (\hat{\sigma}_{A,m}^+ \hat{\sigma}_{A,m+1}^- - {\rm H.c.}) \rangle_{\rm ss}}$. We numerically find that the current scaling with system size to be the same as in 1D spin chains driven by incoherent pumping \cite{Prosen2011-uf,Prosen2015-vk,Bertini2021-px}: ballistic $\langle j \rangle\sim N^0$ in the easy-axis regime $\Delta/J<1$, insulating $\langle j \rangle\sim \exp(-\alpha N)$ for $\Delta/J>1$, and sub-diffusive $j\sim N^{-2}$ at the Heisenberg point $\Delta=1$. These scaling laws are shown in Fig.~\ref{fig-app:transport}. The current is ballistic for $\Delta/J<1$ in both phases $(1)$ and $(2)$ [cf. Fig.~2 of the main text]. It is $\Omega$-dependent, and its non-analytic behavior similar to that of the average magnetization is numerically observed at the critical drive $\Omega_c$.
Finally, we note that although phases $(1)$ and $(3)$ are asymptotically connected at $\Omega=0$, with each approaching the ferromagnetic state, they remain distinct in their transport properties.

\section{Hidden Time Reversal Symmetry}

In this section, derive the Onsager symmetry mentioned in the main text. As a consequence of hidden time reversal symmetry, there are constraints on observables, including that a particular subset of two-time correlation functions obey Onsager time symmetry \cite{Roberts2021-cz}:
\begin{equation}
    {\rm Tr} (\hat{X}(t) \mathcal{J}[\hat{Y}] \hat{\rho} _{\rm ss}) = {\rm Tr} (\hat{Y}(t) \mathcal{J}[\hat{X}] \hat{\rho} _{\rm ss}).
\end{equation}
Here $\mathcal{J}[\hat{O}]  = \hat{\rho}^{1/2}_{\rm ss} \hat T_{\rm h} \hat{O}^\dagger \hat T_{\rm h}^{-1} \hat{\rho}^{-1/2}_{\rm ss} $ is the time reversal superoperator with the hidden time reversal antiunitary operator $\hat T_{\rm h}$. This symmetry is nonlinear in the steady state density matrix, so the time-reversal of generic observables is not necessarily an easily experimentally measurable quantity. However, there is a class of operators that are invariant under time reversal, $\mathcal{J}[\hat X]=\hat X$, hence obey standard Onsager symmetry
\begin{equation}
    \langle \hat{X} (t)  \hat{Y} (0)  \rangle_{\rm ss} = \langle  \hat{Y} (t) \hat{X} (0) \rangle_{\rm ss}.
\end{equation}
This subset always include the effective non-Hermitian Hamiltonian $\hat{H}_{\rm eff}=\hat H - \frac{i}{2}\sum_k \hat L_k$ and the jump operators $\hat L_k$. One may generate several other time-reversal symmetric operators by taking linear combinations and products of $\hat H_{\rm eff}$ and $\hat L_k$ and using the property $\mathcal{J}[\hat{X} \hat{Y}] = \mathcal{J}[\hat{Y}] \mathcal{J}[\hat{X}]$. This implies, for example, that the commutator is anti-symmetric, $\mathcal{J}\{[\hat X,\hat Y]\} = -[\hat X, \hat Y]$, and the nested commutator is symmetric, $\mathcal{J}\{[[\hat X, \hat Y],\hat Y]]\} = [[\hat X, \hat Y],\hat Y]]$.

\begin{figure}[h]
    \centering
    \includegraphics{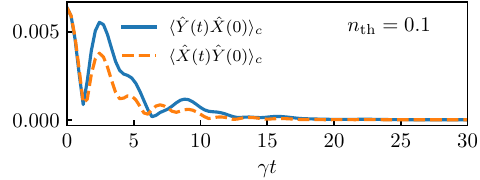}
    \caption{The Onsager symmetry for two-time correlation functions $\hat{X} = \hat{\sigma}_1^-$ and $\hat{\sigma}_1^-  \hat{\sigma}_2^-$ [cf.~Fig.~4(b) in the main text] is broken when the dissipative bath is at finite temperature, here $\Bar{n}_{\rm th} = 0.1$.  This is shown for $N=3$, $\Omega/J = 0.4$, $\Delta/J=0.2$ and $\gamma/J = 1$.}
    \label{fig-app:nth}
\end{figure}

In the case of the coherently driven model, the non-Hermitian Hamiltonian is $\hat{H}_{\rm eff} = \hat{H}_{\rm XXZ}+ \frac{\Omega}{2}\hat{\sigma}^x_N+ \frac{i\gamma}{2} \hat{\sigma}_1^+ \hat{\sigma}_1^- $, and the single jump operator is $\hat\sigma_1^-$.
The latter can be readily recovered through measurements of $\hat \sigma^x_1$ and $\hat\sigma^y_1$. Taking the nested commutator of $\hat H_{\rm eff}$ with $\hat \sigma_1^-$, which is time-reversal symmetric, we find
\begin{equation}
    \mathcal{J} \left[- \frac{1}{2}[[\hat{H}_{\rm eff}, \hat{\sigma}_1^-],\hat{\sigma}_1^-] \right]= - \frac{1}{2}[[\hat{H}_{\rm eff}, \hat{\sigma}_1^-],\hat{\sigma}_1^-] =\hat{\sigma}_1^- \hat{H}_{\rm eff} \hat{\sigma}_1^- = J \hat{\sigma}_{1}^- \hat{\sigma}_{2}^-.
\end{equation}
The expectation value of this operator can be recovered from joint $X$ and $Y$ measurements of the two spins.
Thus, the correlation function $\langle \hat\sigma_1^-(t)(\hat \sigma_1^-\sigma_2^-)(0)\rangle$ obeys Onsager symmetry, as noted in the main text:
\begin{equation}
    \langle \hat{\sigma}_1^- (t)  (\hat{\sigma}_1^-  \hat{\sigma}_2^-) (0)  \rangle_{\rm c} = \langle   (\hat{\sigma}_1^-  \hat{\sigma}_2^-) (t) \hat{\sigma}_1^- (0) \rangle_{\rm c}. \label{eq-app:Onsager}
\end{equation}
Hidden time reversal symmetry can be frequently broken by either Hamiltonian or dissipative perturbations. For example, consider when the dissipative bath is at finite temperature, with thermal occupation $\Bar{n}_{\rm th} \neq 0$. The non-zero thermal occupation modifies Eq.~\eqref{eq-app:sysAqme} to $\partial_t \hat{\rho} = \hat{\mathcal{L}} \hat{\rho} = -i[\hat{H}_{\rm XXZ}+ \frac{\Omega}{2} \hat{\sigma}_N^x, \hat{\rho}] + \Bar{n}_{\rm th}\gamma \mathcal{D}[\hat{\sigma}_1^+] \hat{\rho}. + (1+\Bar{n}_{\rm th} )\gamma \mathcal{D}[\hat{\sigma}_1^-] \hat{\rho}$. In this case, the Onsager symmetry of the correlation functions in \cref{eq-app:Onsager} is broken [See \cref{fig-app:nth}].

\section{CQA solution for the incoherent model}
\label{sec-app:cqa-incoherent}
As mentioned in the main text, the CQA method is also applicable to other previously solved models \cite{Karevski2013-ju,Prosen2011-uf,Prosen2014-od}. Here we demonstrate the CQA solution to a well known exact solution: the incoherent pumping and loss XXZ chain solved in Ref.~\cite{Prosen2011-uf}. The system is described by the Lindblad master equation 
\begin{equation}
    \partial_t \hat{\rho} = \mathcal{L}\hat{\rho} = -i[\hat H_{\rm XXZ}, \hat{\rho}] + \gamma \mathcal{D}[\hat{\sigma}_1^-] \hat{\rho}+ \gamma \mathcal{D}[\hat{\sigma}_N^+] \hat{\rho}, \label{eq-app:incoherent-me}
\end{equation} 
where the Hamiltonian is the usual $\hat{H}_{\rm XXZ} = \sum_{j = 1}^{N-1} (J \hat{\sigma}_j^x  \hat{\sigma}_{j+1}^x + J \hat{\sigma}_j^y  \hat{\sigma}_{j+1}^y  + \Delta \hat{\sigma}_j^z\hat{\sigma}_{j+1}^z)$ with loss on site 1 and incoherent pumping on site $N$. We take the pumping and loss rates to be equal.
we make the CQA ansatz by constructing a mirrored absorber, $\hat H^{(B)} = -\hat H^{(A)}$, and coupling the system and absorber via two chiral waveguides, yielding the CQA master equation
\begin{align}
    \partial_t \hat{\rho}_{\rm AB} &= \mathcal{L}_{\rm AB} \hat{\rho}_{\rm AB} = - i[\hat H^{(AB)}, \hat{\rho}_{\rm AB}] + \gamma \mathcal{D}[\hat{L}_{\rm AB}^{(1)}] \hat{\rho}_{\rm AB}+ \gamma \mathcal{D}[\hat{L}_{\rm AB}^{(2)}] \hat{\rho}_{\rm AB}. \label{eq-app:incoherent-double}\\
    \hat H^{(AB)} &= \hat{H}^{(A)}_{\rm XXZ} - \hat{H}^{(B)}_{\rm XXZ} + \hat{H}_{\rm c,1}+ \hat{H}_{\rm c,N}, \\
    \hat{H}_{\rm c,1} &= -\frac{i\gamma}{2}(\hat \sigma_{A,1}^+\hat\sigma_{B,1}^- - {\rm h.c.}),\quad \hat{H}_{\rm c,N} = -\frac{i\gamma}{2}(\hat \sigma_{A,N}^-\hat\sigma_{B,N}^+ - {\rm h.c.}).
\end{align}
Here $\hat L_{AB}^{(1)} = \hat\sigma_{A,1}^- - \hat\sigma_{B,1}^-$ and $\hat L_{AB}^{(2)} = \hat\sigma_{A,N}^+ - \hat\sigma_{B,N}^+$ are the collective dissipators mediated by the chiral waveguides.

A pure steady state $\ket{\psi_{\rm CQA}^I}$ must satisfy the dark state conditions
\begin{equation}
    \hat{H}_{\rm AB}\ket{\psi_{\rm CQA}^I} = 0, \quad \hat{L}^{(1)}_{\rm AB}\ket{\psi_{\rm CQA}^I} = 0, \quad \hat{L}^{(2)}_{\rm AB}\ket{\psi_{\rm CQA}^I} = 0.
\end{equation}
Notice that \cref{eq-app:incoherent-me} and \cref{eq-app:master} are identical except for the driving term on site $N$: $\hat H_{\rm d}$ in \cref{eq-app:incoherent-me} is replaced by $\hat H_{{\rm c},N}$.
We make the same MPS ansatz $|\psi^I_{\rm CQA}\rangle = \sum_{s_1,\cdots, s_N} \vb{v_L^{\dagger}} \hat{A}_{s_1} \cdots \hat{A}_{s_N} \vb{v_R} |s_1 \cdots s_N \rangle$ as before. We do not assume these to be the same matrices and vectors as for the coherently driven model, but the Hamiltonian dark state condition in the bulk and on the loss site $1$ can be satisfied by constructing the same matrices as above [cf.~Eq.~\eqref{eq-app:definite-A}], and by taking $\vb{v_L^\dagger}=\langle 0|$. Moreover, the dark state condition for the pumping on site $N$ imposes the constraint $\hat{A}_0 \vb{v_R} = 0$, which is satisfied by $\vb{v_R}=|0\rangle$. Note that the boundary vectors being identical preserves the $\mathbb{Z}_2$ symmetry of the master equation. 

We, therefore, find the CQA solution of the incoherently driven model to be
\begin{equation}
    |\psi^I_{\rm CQA}\rangle = \sum_{s_1,\cdots, s_N} \bra{0}\hat{A}_{s_1} \cdots \hat{A}_{s_N}\ket{0} |s_1 \cdots s_N \rangle,
\end{equation}
where $s_i \in \{0_i,1_i,T_i\}$, and the MPS matrix coefficients are given by Eqs.~\eqref{eq-app:a-analytical} and \eqref{eq-app:b-analytical}. One may make the gauge choice $c_k = 1$, hence Eq.~\eqref{eq-app:b-analytical} gives the coefficients $b_k$.
We note that our CQA solution is a specific purification of the exact solution density matrix found in Ref.~\cite{Prosen2011-uf}; moreover, we may follow the same steps as above and trace out the absorber system to write $\hat \rho_{\rm ss}^I = \hat \Psi_I \hat{\Psi}_I^\dagger$ as a Cholesky decomposition. Finally we point out that $|\psi_{\rm CQA}\rangle$ is a pure \emph{current-carrying} steady state of \cref{eq-app:incoherent-double}.

\bibliography{xxz-transport}